\documentclass[12pt]{iopart}
\usepackage{iopams}
\usepackage{epsfig}
\newtheorem{theorem}{Theorem}

\newtheorem{lemma}[theorem]{Lemma}

\newtheorem{remark}[theorem]{Remark}

\newenvironment{proof}[1][Proof]{\noindent\textbf{#1.} }{\ \rule{0.5em}{0.5em}}

\begin{document}


\title[Determinant solution for the TASEP with parallel update II. Ring geometry]{Determinant solution for the Totally Asymmetric Exclusion Process
with parallel update II. Ring geometry.}
\author{A.M. Povolotsky$^{1,2,\dag}$}
\author{V.B. Priezzhev$^{2,\ddag}$}
\address{$^1$School of Theoretical Physics, Dublin Institute for Advanced Studies,
Dublin, Ireland, \\
$^{2}$Bogoliubov Laboratory of Theoretical Physics, Joint Institute for
Nuclear Research, 141980 Dubna, Russia}
\eads{$^\dag$\mailto{alexander.povolotsky@gmail.com},$^\ddag$\mailto{priezzvb@thsun1.jinr.ru}}

\begin{abstract}
Using the Bethe ansatz we obtain the determinant expression for the time dependent transition
probabilities in the totally asymmetric exclusion process with parallel
update on a ring. Developing a
method of summation over the roots of Bethe equations based on the
multidimensional analogue of the Cauchy residue theorem, we construct the
resolution of the identity operator, which allows us to calculate the
matrix elements of the evolution operator and its powers. Representation of
results in the form of an infinite series elucidates connection
to other results obtained for the ring geometry. As a byproduct we also obtain
the generating function of the joint probability distribution of
 particle configurations and the total distance traveled by the particles.
\end{abstract}

\pacs{05.40.+j, 02.50.-r, 82.20.-w}
\maketitle
%

\section{Introduction}

The present paper can be viewed as a continuation of the paper \cite%
{povolotsky priezzhev}, where the transition probabilities for the
totally asymmetric simple exclusion process (TASEP) with parallel update have been obtained
for the 1D infinite lattice, generalizing well-known result of Sch\"{u}tz \cite{schutz} for
the continuous time TASEP. The analytic method,
developed in the first half of that paper, was based on the use of Bethe
eigenvectors of the evolution operator obtained in \cite{povolotsky mendes}.
The problem was to find a continuous spectrum, i.e. the integration measure,
which would allow one to construct the solutions of the master equation out
of the Bethe eigenvectors, and in particular would give a resolution of the
identity operator in the integral form. The proof of the formula for the
resolution of the identity operator, was the main technical result of the
paper which yielded the final determinant formula for the transition
probability.

A peculiarity of the finite ring is that the spectrum
becomes discrete due to the periodic boundary conditions, being defined by
the system of the algebraic Bethe equations (BE). These equations usually
cannot be solved exactly. Most attempts to extract any information from
them are related to the thermodynamic limit, where they
can be reinterpreted in terms of a single integral equation. For the finite
lattices, not only the exact form of the spectrum but even the issue of its
completeness are far from being well understood for most of the integrable
models.

Fortunately, for the TASEP the
situation is a bit better due to a very special factorization property
possessed by the BE. This property was observed already in one of the first
works on the continuous time TASEP by Gwa and Spohn \cite{gwa spohn}, where
it was used to get the asymptotical behaviour of the spectral gap. One of
the most impressive results exploiting the special structure of the BE in
TASEP was the exact derivation of the largest eigenvalue of the equation for
the generating function of the integrated particle current, due to Derrida
and Lebowitz \cite{derrida lebowitz}. Using the Cauchy residue theorem, they
managed to evaluate explicitly the sum over roots of the BE for the
particular solution corresponding to the groundstate of the evolution
operator. Recently, some peculiar details of the spectrum structure for the
continuous time TASEP have been also studied in \cite{golinelli mallick}.

The solution of the Cauchy problem, i.e. finding the solution, given the
initial conditions, for the master equation for the TASEP on the ring in
continuous time, and also with backward ordered update, has been recently
proposed by one of the authors \cite{priezzhev_preprint}. To our
knowledge, this is still the only example of full solution to the Cauchy
problem for an integrable model that cannot be reduced to free fermions.
The method of the solution is based on a geometric approach to the Bethe
ansatz (BA), which treats trajectories of the interacting particles as
free, noninteracting, but supplied with the additional statistical
weights in such a way that the interaction is taken into account. The
idea of the solution on a ring was to represent it as an infinite line
with periodic patterns of synchronously moving interacting particles. As in
the geometric formulation one works only with the ensembles of particle
trajectories of a finite length, the issues related to the structure of the
eigenspace of the evolution operator can be by-passed. Motivated by the
geometric solution, its analytic counterpart called the detailed BA has
been also proposed \cite{priezzhev}. Being defined on the infinite
lattice, the detailed BA has a form of the infinite formal sum, and can be
considered as a generating function of particle trajectories. As a result,
the problem becomes infinite lattice-like and one can again use
integration over the continuum to reconstruct the transition probabilities.
The mathematical meaning of the term-by-term integration of the infinite
formal series was not yet well understood, \ and, as we will see below, it
is just equivalent to summation over the discrete spectrum
given by the BE.

The geometric method was applied to the TASEP with parallel update on the
infinite lattice in the second half of our first paper \cite{povolotsky
priezzhev}. Its generalization to the ring geometry follows in line with the
continuous time version and as such is straightforward. Therefore, the aim
of this paper is not only to obtain the solution of the Cauchy problem for
the TASEP with parallel update, but also to establish a bridge between
existing solutions on a ring and the standard BA techniques. Namely, we show
that the idea of Derrida and Lebowitz exploiting the Cauchy residue theorem
to obtain the roots of the BE corresponding to the groundstate solution can be
developed much further and used to perform an exact summation over the
whole spectrum of solutions. In such a way we obtain the integral
representation for the resolution of the identity operator and, as a
consequence, for the solution of the Cauchy problem for the master equation
as well. Then we show that the expression under the integral can be expanded
into a uniformly converging power series, equivalent to the detailed BA,
which being integrated term by term yields finally the multiple infinite sums
coming from the geometric solution and from the detailed BA.

The paper is organized as follows. In section II, we formulate the
dynamical rules of the model and announce the final result. In section III,
we describe the BA for the infinite lattice and give the BE for the ring. In section
IV, we discuss the details of the analytic structure of the BE solutions.
In section V we develop a technique of evaluation of sums over the
Bethe roots. In section VI, we prove a formula for the resolution of the
identity operator and obtain a formula for transition probabilities in
the form of infinite sums. Then, to show that the sums are actually finite, we
estimate the number of nonzero terms they contain. In the last section VII,
we discuss the results obtained and subsequent perspectives.

\section{Formulation of the model and results}
We consider $P$ particles on the 1D ring consisting of $L$ sites.
 The notation
\begin{equation}
\rho =P/L
\end{equation}%
is used for the density of particles.
The model has totally asymmetric dynamics, i.e. all particles
jump only in one direction, which we refer to as forward.
At each step of discrete time $t$, a particle from each occupied site,

\begin{description}
\item[i] \textit{takes one step forward with probability }$v$
\end{description}

or

\begin{description}
\item[ii] \textit{stays with probability }$(1-v)$,
\end{description}
provided that the target site is vacant. When the next site is occupied,

\begin{description}
\item[iii] \textit{a particle stays with probability }$1$.
\end{description}

All sites are updated simultaneously. We define a configuration $X$ of
particles by the set of their coordinates $X=\left\{ x_{1},x_{2},\ldots
,x_{P}\right\} $, written in strictly increasing order,
\begin{equation}
x_{1}<x_{2}<\cdots <x_{P}.  \label{x_1<x_2<...<x_M.}
\end{equation}%
The finite ring geometry implies also that the coordinates are confined to
the values from $1$ to $L$.
\begin{equation}
x_{1}\geq 1,x_{P}\leq L  \label{Ring limits}
\end{equation}%
The probability $P_{t}(X)$ for the system to be in a configuration $X$ at
time $t$ obeys the Markov equation%
\begin{equation}
P_{t+1}(X)=\sum_{\left\{ X^{\prime }\right\} }T(X,X^{\prime
})P_{t}(X^{\prime }),  \label{master}
\end{equation}%
where $T(X,X^{\prime })$ is the probability of the transition from $%
X^{\prime }$ to $X$ for one time step. The transition probability $%
T(X,X^{\prime })$ defined by the above dynamical rules is a product of
factors, each corresponding to a particular cluster of particles in the
initial configuration $X^{\prime }$. The word "cluster" refers to a group of
particles, which has no empty sites between occupied sites and two empty sites at
the ends. The value of these factors is either $v$\ or $(1-v)$\ depending on
whether or not the first particle of a given cluster jumps during the
transition from $X^{\prime }$\ to $X.$%
\begin{equation}
T(X,X^{\prime })=\prod\limits_{i=1}^{\mathcal{N}_{c}\left( X^{\prime
}\right) }\left( 1-v\right) ^{1-m_{i}}v^{m_{i}}
\label{transition probability}
\end{equation}%
Here $m_{i}=0,1$ is the number of particles hopping from $i$-th cluster of $%
X^{\prime }$, and ${\mathcal{N}_{c}\left( X^{\prime }\right)}$ is the number
of clusters in $X^{\prime }$.

It has been shown in \cite{povolotsky priezzhev} that in the case of the
infinite lattice, when  constraint (\ref{Ring limits}) is omitted,
the conditional probability, $P(X;t|X^{0};0)$%
, for the system to be in a configuration $X=\left\{ x_{1},x_{2},\ldots
,x_{P}\right\} $ at time $t$, given it was in a configuration $X^{0}=\left\{
x_{1}^{0},x_{2}^{0},\ldots ,x_{P}^{0}\right\} $ at time $0$, is the
following quotient
\begin{equation}
P(X;t|X^{0};0)=\frac{\mathcal{F}_{\infty }\left( X,X^{0},t\right) }{\mathcal{%
F}_{\infty }\left( X,X,0\right) }\mathbf{.}  \label{P(X,t|X_0,0)}
\end{equation}%
The function $\mathcal{F}_{\infty }\left( X,Y,t\right)$ depending on two
particle configurations $X=\left\{ x_{1},x_{2},\ldots ,x_{P}\right\} $ and $%
Y=\left\{ y_{1},y_{2},\ldots ,y_{P}\right\} $ and on time $t$ is given by
the determinant of $P\times P$ matrix
\begin{equation}
\mathcal{F}_{\infty }\left( X,Y,t\right) =\det \left[ f(i-j,x_{i}-y_{j},t)%
\right] _{1\leq i,j\leq P},  \label{F_infinity}
\end{equation}%
where the matrix elements are defined in terms of a single function $f\left(
a,b,t\right) $ expressed via the Gauss hypergeometric functions:

\begin{equation}
\hspace{-15mm}~f(a,b,t)=\left( 1-v\right) ^{t}%
\cases{
\begin{array}{ll}
\left( \frac{v}{v-1}\right) ^{b}\frac{(-t-a)_{b}}{b!}\left. _{2}F_{1}\right.
\left(
\begin{array}{c}
a,-t-a+b \\
b+1%
\end{array}%
;\frac{v}{v-1}\right) & b>0 \\
\frac{\left( a\right) _{-b}}{\left( -b\right) !}\left. _{2}F_{1}\right.
\left(
\begin{array}{c}
a-b,-t-a \\
-b+1%
\end{array}%
;\frac{v}{v-1}\right) & b\leq 0%
\end{array}%
.%
}
\label{f(a,b,t)}
\end{equation}%
The notation $\left( a\right) _{n}$ is for the shifted factorial $\left(
a\right) _{n}=a(a+1)\cdots \left( a+n-1\right) $.

The aim of the present article is to show that on the ring of size $L$
the same quantity $P(X;t|X^{0};0)$ is also given by similar quotient of two
terms, both expressed as a single function $\mathcal{F}_{L}\left(
X,Y,t\right) $ of particle configurations $X,Y$ and time $t$,
\begin{equation}
P(X;t|X^{0};0)=\frac{\mathcal{F}_{L}\left( X,X^{0},t\right) }{\mathcal{F}%
_{L}\left( X,X,0\right) }  \label{F_ring}
\end{equation}%
with the arguments in the numerator and the denominator taken as in (\ref%
{P(X,t|X_0,0)}). However, for the finite ring, the function $%
\mathcal{F}_{L}\left( X,Y,t\right) $ is the $P$-tuple sum of
determinants
\begin{eqnarray}
&\mathcal{F}_{L}\left( X,Y,t\right) =\sum\limits_{n_{1}=-\infty }^{\infty
}\cdots \sum\limits_{n_{P}=-\infty }^{\infty }\left( -1\right)
^{(P-1)\sum_{i=1}^{P}n_{i}}  \nonumber \\
&\times \det \left[ f\left(
i-j+Pn_{i}-\sum\limits_{k=1}^{P}n_{k},x_{i}-y_{j}+n_{i}L,t\right) \right]
_{1\leq i,j\leq P},  \label{F(X,Y,t)_ring}
\end{eqnarray}%
unlike the single determinant in the case of infinite lattice. The matrix
elements of the corresponding matrices are still given in terms of the
function $f\left( a,b,t\right) $ defined in (\ref{f(a,b,t)}), but its
arguments depend now  not only on the matrix indices $i,j$ but also on
summation indices $n_{k}$. Below we argue that though formally the sums are
infinite, they contain only a finite number of nonzero terms for any finite time $%
t$. Furthermore, the denominators of (\ref{P(X,t|X_0,0)}) and (\ref%
{F_ring}) depend only on the number of clusters in the corresponding
configuration $X$ being equal to
\begin{equation}
\mathcal{F}_{\infty }\left( X,X,0\right) =\mathcal{F}_{L}\left( X,X,0\right)
=\left( 1-v\right) ^{\mathcal{N}_{c}\left( X\right) -P}
\label{F(X,X,0)}
\end{equation}

\begin{remark}
One has to define $\mathcal{N}_{c}\left( X\right) $ in (\ref{F(X,X,0)})
as a function of $X$ separately for the infinite and finite lattices.
Indeed, the particles, which occupy the sites $1$ and $L$ on the ring,
belong to the same cluster, while on the infinite lattice they do not.
Therefore, the value of $\mathcal{N}_{c}(X)$ can be different for these two
cases, even though the coordinates of particles formally coincide.
\end{remark}

\section{Bethe ansatz.\label{Sec Bethe ansatz}}

\subsection{Infinite lattice.}

We first remind the reader of the technique used to deal with the infinite
lattice case \cite{povolotsky priezzhev}. Consider $P$ particles on the infinite
lattice. The particle configurations are given by $P-$tuples of particle
coordinates which are unbounded integers being selected from the set

\begin{equation}
\mathbb{Z}_{<}^{P}\equiv \left\{ X\in \mathbb{Z}^{P};x_{1}<x_{2}<\cdots
<x_{P}\right\} .
\end{equation}%
Let us introduce the infinite dimensional vector space $V_{\infty }$ over
the field of complex numbers $\mathbb{C}$ given by the linear span of the
basis
\begin{equation}
\mathcal{X}_{\infty }=\left\{ \left\vert X\right\rangle :X\in \mathbb{Z}%
_{<}^{P}\right\} ,
\end{equation}%
i.e. the set of the vectors labelled by the particle configurations. In
addition one introduces the basis of the dual space $V_{\infty }^{\ast }$,
which is the span of the dual basis,
\begin{equation}
\mathcal{X}_{\infty }^{\ast }=\left\{ \left\langle X\right\vert :X\in
\mathbb{Z}_{<}^{P}\right\},
\end{equation}%
with the inner product defined by
\begin{equation}
\left\langle X|X^{\prime }\right\rangle =\delta (X,X^{\prime }).
\label{<X|X'>=delta(X,X')}
\end{equation}%
Below the bases $\mathcal{X}$,$\mathcal{X}^{\ast }$ will be referred to as
the configurational left and right bases respectively, unlike the left and
right eigenbases of the evolution operator to be considered. The evolution
operator $\mathbf{T}$ is defined in terms of the transition probabilities $%
T(X,X^{\prime })$ defined in (\ref{transition probability})%
\begin{equation}
\mathbf{T=}\sum_{X\in \mathcal{X}_{\infty },X^{\prime }\in \mathcal{X}%
_{\infty }^{\ast }}\left\vert X\right\rangle T(X,X^{\prime })\left\langle
X^{\prime }\right\vert .  \label{T}
\end{equation}%
The problem under consideration is to find the transition probability $%
P(X;t|X^{0};0)$ from a configuration $X^{0}$ to $X$ for $t$ steps, which is
nothing but the corresponding matrix element of the operator $\mathbf{T}^{t}$
\begin{equation*}
P(X;t|X^{0};0)=\left\langle X|\mathbf{T}^{t}|X_{0}\right\rangle .
\end{equation*}

\begin{figure}[tbp]
\unitlength=1mm {\makebox(160,100)[cc]{\psfig{file=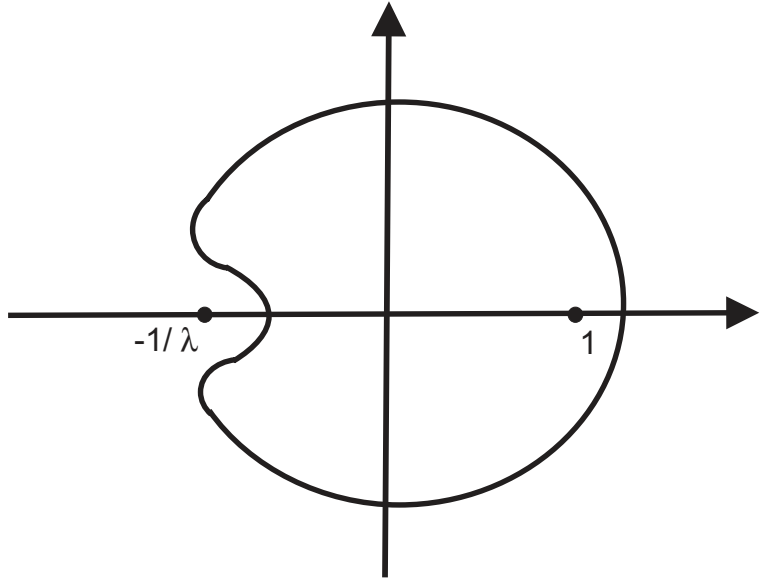,width=100mm}}}
\caption{Shape of the contour $\Gamma ^{\infty }$. It must encircle
 the points $z=1$ and $z=0$ while the point $z=-1/\lambda$ stays outside. }
\label{contour}
\end{figure}

It was shown in \cite{povolotsky mendes}, \cite{povolotsky priezzhev}
that the evolution operator $\mathbf{T}$ has the left and right
eigenvectors $\left\vert B_{Z}\right\rangle $, $\left\langle \overline{B}%
_{Z}\right\vert $ parametrized by $P$-tuple complex parameter
\begin{equation*}
Z\equiv \left\{ z_{1},\ldots ,z_{P}\right\} \in \mathbb{C}^{P}.
\end{equation*}%
The eigenvectors corresponding to the same value of $Z$ solve the left and
right eigenproblems,
\begin{equation}
\mathbf{T}\left\vert B_{Z}\right\rangle =\Lambda (Z)\left\vert
B_{Z}\right\rangle ,\left\langle \overline{B}_{Z}\right\vert \mathbf{T}%
=\Lambda (Z)\left\langle \overline{B}_{Z}\right\vert ,  \label{eigenproblem}
\end{equation}%
associated with the same eigenvalue
\begin{equation}
\Lambda \left( Z\right) =\left( 1+\lambda \right)
^{-P}\prod\limits_{i=1}^{P}\left( 1+\lambda z_{i}\right) ,  \label{Lambda(Z)}
\end{equation}%
where we introduce the parameter $\lambda $
\begin{equation}
\lambda =\frac{v}{1-v}.  \label{lambda}
\end{equation}%
The projection $\left\langle X|B_{Z}\right\rangle $ of the right eigenvector
$\left\vert B_{Z}\right\rangle $ to the configuration $X$ is given by the Bethe ansatz
\begin{equation}
\left\langle x_{1},\ldots ,x_{P}|B_{Z}\right\rangle =W(X)\sum_{\left\{
\sigma \right\} }A_{\sigma _{1}\ldots \sigma _{P}}z_{\sigma
_{1}}^{-x_{1}}\ldots z_{\sigma _{P}}^{-x_{P}},  \label{Bethe ansatz_right}
\end{equation}%
supplied with the additional configuration-dependent factor $W(X)$. The
latter is proportional to the stationary measure of the configuration $X$
and is defined in terms of the number of clusters $\mathcal{N}_{c}\left(
X\right) $ in the configuration $X$
\begin{equation}
W\left( X\right) =\left( 1+\lambda \right) ^{\mathcal{N}_{c}\left( X\right)
-P}.  \label{W(X)}
\end{equation}%
The amplitudes $A_{\sigma }$, are indexed by the permutations $\sigma =$ $%
\left\{ \sigma _{1},\cdots ,\sigma _{P}\right\} $ of the integers $1,\ldots
,P$. An elementary transposition of two indices $(ij)$ results in the
amplitude $A_{\sigma }$ being multiplied by the scattering factor $-S\left(
z_{i},z_{j}\right) $,%
\begin{equation}
A_{\ldots ij\ldots }=-S\left( z_{i},z_{j}\right) A_{\ldots ji\ldots },
\label{A=SA}
\end{equation}%
of the following form
\begin{equation}
S\left( z_{i},z_{j}\right) \equiv \frac{1-1/z_{i}}{1-1/z_{j}}\frac{1+\lambda
z_{j}}{1+\lambda z_{i}}.  \label{S}
\end{equation}%
A remarkable property of the scattering factor $S\left( z_{i},z_{j}\right)
$ is that it is a product of two factors, each being dependent only on one
of the two parameters $z_{i}$, $z_{j}$. This property allows one to
represent the amplitude $A_{\sigma }$ in a simple product form
\begin{equation}
A_{\sigma _{1}\ldots \sigma _{P}}=\left( -1\right) ^{\left\vert \sigma
\right\vert }\prod\limits_{i=1}^{P}\left( \frac{1+\lambda z_{\sigma _{i}}}{%
1-1/z_{\sigma _{i}}}\right) ^{i-\sigma _{i}},  \label{A_sigma}
\end{equation}%
where each multiple depends only on one of the parameters $z_{1},\ldots
,z_{P}$. Unlike the right eigenvector, the left one, $\left\langle \overline{%
B}_{Z}\right\vert ,$ has no factor $W(X)$, while the Bethe part can be
obtained from that of (\ref{Bethe ansatz_right}) by the change $A_{\sigma
}\rightarrow A_{\sigma }^{-1}$, $x_{i}\rightarrow -x_{i}$.%
\begin{equation}
\left\langle \overline{B}_{Z}|x_{1},\ldots ,x_{P}\right\rangle
=\sum_{\left\{ \sigma \right\} }A_{\sigma _{1}\ldots \sigma
_{P}}^{-1}z_{\sigma _{1}}^{x_{1}}\ldots z_{\sigma _{P}}^{x_{P}}.
\label{bethe ansatz left}
\end{equation}%
Then, that the projection of the eigenvectors $\left\langle
\overline{B}_{Z}|X\right\rangle $ and $\left\langle X|B_{Z}\right\rangle $
to the basis vectors corresponding to a particular configuration $X$ can be
represented in the form of determinants%
\begin{eqnarray}
\left\langle X|B_{Z}\right\rangle &=&W(X)\det \mathbf{B,}  \label{<X|B>} \\
\left\langle \overline{B}_{Z}|X\right\rangle &=&\det \overline{\mathbf{B}},
\label{<B|X>}
\end{eqnarray}%
where the matrix elements $B_{ij}$ and $\overline{B}_{ij}$
are given by%
\begin{equation}
B_{ij}=1/\overline{B}_{ij}=\left( \frac{1+\lambda z_{j}}{1-1/z_{j}}\right)
^{i-j}z_{j}^{-x_{i}}.  \label{B_ij}
\end{equation}%
One of the main results of \ \cite{povolotsky priezzhev} is the proof of the
formula for the resolution of the identity operator,
\begin{equation}
\int \left\vert B_{Z}\right\rangle \left\langle \overline{B}_{Z}\right\vert
d\mu (Z)=\mathbf{E},  \label{completeness}
\end{equation}%
where $\mathbf{E}$ is the identity operator, $d\mu (Z)=\left( P!\right)
^{-1}\prod\nolimits_{i=1}^{P}\left( dz_{i}/2\pi \mathrm{i}z_{i}\right) $ is
the integration measure, and the integration is performed independently\
over each $z_{i}$ along the contour $\Gamma ^{\infty }$ encircling the
points $z=0$ and $z=1$, while the point $z=-1/\lambda $ stays outside (figure %
\ref{contour}). Practically the proof was given by the direct evaluation of
the integral in the configurational basis,%
\begin{equation}
\int \left\langle X^{\prime }|B_{Z}\right\rangle \left\langle \overline{B}%
_{Z}|X\right\rangle d\mu (Z)=\delta _{X,X^{\prime }}.
\label{completeness(basis)}
\end{equation}%
As soon as the equality (\ref{completeness}) is established, one can insert
the identity operator into the matrix elements,
\begin{eqnarray}
\left\langle X\right\vert \mathbf{T}^{t}\left\vert X^{0}\right\rangle
&=&\int \left\langle X|\mathbf{T}^{t}|B_{Z}\right\rangle \left\langle
\overline{B}_{Z}|X^{0}\right\rangle d\mu (Z) \\
&=&\int \Lambda ^{t}\left( Z\right) \left\langle X|B_{Z}\right\rangle
\left\langle \overline{B}_{Z}|X^{0}\right\rangle d\mu (Z)
\end{eqnarray}%
so that the final result (\ref{P(X,t|X_0,0)}-\ref{f(a,b,t)}) immediately
follows from the explicit form of the eigenvalues and the eigenvectors.

One of the observations made in \cite{povolotsky priezzhev} was that there is no
any obvious procedure of choosing a contour and a measure of integration.
However, validity of a particular choice can be verified a posteriori by
proving the resolution of the identity operator, which is fulfilled by a
direct evaluation of l.h.s. of (\ref{completeness(basis)}). It will be
clear below that the form of the contour can be validated directly by
considering the infinite lattice as a limiting case of the ring of the size $%
L$, which grows to infinity. Then, the $Z-$spectrum of the model is obtained
as a continuous limit of the discrete spectrum for the finite
system.

\subsection{The ring.\protect\bigskip}

What does change when one confines the system to the ring of finite length $%
L $? Apparently, the total number of particle configurations becomes finite.
The same must be true for the dimension of vector space $V_{L}$ defined as the span of the
basis
\begin{equation}
\mathcal{X}_{L}=\left\{ \left\vert X\right\rangle :X\in \mathbb{Z}%
_{<,L}^{P}\right\} ,
\end{equation}%
where $\mathbb{Z}_{<,L}^{P}$ is the domain for particle coordinates on the
ring.
\begin{equation}
\mathbb{Z}_{<,L}^{P}\equiv \left\{ X\in \mathbb{Z}^{P};1\leq
x_{1}<x_{2}<\cdots <x_{P}\leq L\right\} .  \label{Z^P_<,L}
\end{equation}%
As before, the dual space $V_{L}^{\ast }$ is spanned by the dual basis
\begin{equation}
\mathcal{X}_{L}^{\ast }=\left\{ \left\langle X\right\vert :X\in \mathbb{Z}%
_{<,L}^{P}\right\} ,
\end{equation}%
with the inner product defined by
\begin{equation*}
\left\langle X|X^{\prime }\right\rangle =\delta (X,X^{\prime }).
\end{equation*}%
Obviously, the dimensions of the spaces are equal to the total number of the
basis vectors, which is equal to the number of particle configurations:%
\begin{equation}
\dim \left( V_{L}\right) =\dim \left( V_{L}^{\ast }\right) =\left(
\begin{array}{c}
L \\
P%
\end{array}%
\right) .  \label{dim(H_L)}
\end{equation}%
The solution of the eigenproblem turns out to be analogous to the infinite
lattice case. The only though important difference is that, due to the
finite dimension of the vector space, a finite number of  independent
eigenvectors can exist. Technically this follows from the fact that
vectors $\left\vert B_{Z}\right\rangle $ and $\left\langle \overline{B}%
_{Z}\right\vert $ are now the eigenvectors of $\mathbf{T}$ only for a
finite discrete set of the values of the parameter $Z$. This set is to be
defined from the system of algebraic equations, which follow from imposing
the periodic boundary conditions. This is the set one has to
sum over, when constructing the resolution of the identity operator similar
to (\ref{completeness}). Of course the latter is correct provided that this
set is large enough to ensure that the corresponding eigenvectors to form the
complete bases of $V_{L}$ and $V_{L}^{\ast }$

For further convenience we slightly generalize the problem. Consider the
generating function $F_{t}^{\gamma }(X;t|X^{0};0)$ of the joint probability $%
P_{t}(X,J;t|X^{0},0;0)$ for the system to be in a configuration $X$ at time $%
t$, the total distance travelled by particles being $J$, given the
initial configuration $X^{0}$.
\begin{equation}
F_{t}^{\gamma }(X;t|X^{0};0)=\left\langle \exp \left( \gamma y_{t}\right)
\right\rangle _{X}=\sum\limits_{J=0}^{\infty }e^{\gamma
J}P_{t}(X,J;t|X^{0},0;0),
\end{equation}%
The evolution equation for $F_{t}^{\gamma }(X,t|X^{0},0)$ is similar to the
original equation (\ref{master}) for the probability, with the only minor
change: the transition probabilities must be multiplied by the factor $%
e^{\gamma }$ per each jumping particle:
\begin{equation}
T_{\gamma }(X,X^{\prime })=\prod\limits_{i=1}^{\mathcal{N}_{c}\left(
X^{\prime }\right) }\left( 1-v\right) ^{1-m_{i}}\left( e^{\gamma }v\right)
^{m_{i}}.
\end{equation}%
Apparently, the limit $\gamma \rightarrow 0$ restores the original Markov
equation for the probability of a configuration
\begin{equation}
P_{t}(X;t|X^{0};0)=\lim_{\gamma \rightarrow 0}F_{t}^{\gamma }(X;t|X^{0};0).
\end{equation}%
All elements of the above Bethe ansatz technique can be directly extended to the case of
nonzero $\gamma $ yielding a minor change in the eigenvalue (%
\ref{Lambda(Z)})
\begin{equation}
\Lambda _{\gamma }\left( Z\right) =\left( 1+\lambda \right)
^{-P}\prod\limits_{i=1}^{P}\left( 1+e^{\gamma }\lambda z_{i}\right)
\end{equation}%
and in the scattering factor (\ref{S})
\begin{equation}
S^{\gamma }\left( z_{i},z_{j}\right) \equiv \frac{e^{\gamma }-1/z_{i}}{%
e^{\gamma }-1/z_{j}}\frac{1+e^{\gamma }\lambda z_{j}}{1+e^{\gamma }\lambda
z_{i}}
\end{equation}%
which in turn changes the BA amplitudes, (\ref{A_sigma})
\begin{equation}
A_{\sigma _{1}\ldots \sigma _{P}}^{\gamma }=\left( -1\right) ^{\left\vert
\sigma \right\vert }\prod\limits_{i=1}^{P}\left( \frac{1+\lambda z_{\sigma
_{i}}e^{\gamma }}{e^{\gamma }-1/z_{\sigma _{i}}}\right) ^{i-\sigma _{i}}.
\end{equation}%
These amplitudes being substituted to the the BA for the right and left eigenvectors(\ref%
{Bethe ansatz_right},\ref{bethe ansatz left}) yield an expression for the
eigenvectors of $\mathbf{T}_{\gamma }$. All the expressions obtained for the
infinite lattice are valid for the ring geometry, until the coordinates of
particles take values at the boundary of the  domain (\ref{Z^P_<,L}).
For the expressions to be valid  at the boundary as well,
one has to impose periodic boundary
conditions
\begin{equation}
\left\{ x_{1},x_{2},\ldots ,x_{P}\right\} =\left\{ x_{2},\ldots
,x_{P},x_{1}+L\right\} .
\end{equation}%
Applied to the Bethe ansatz (\ref{Bethe ansatz_right},\ref{bethe ansatz left}%
), the boundary conditions yield a system of algebraic Bethe equations (BE)
\begin{equation}
z_{i}^{L}=\left( -1\right) ^{P-1}\prod\limits_{j=1}^{P}\frac{z_{i}}{z_{j}}%
\frac{(z_{j}e^{\gamma }-1)}{(z_{i}e^{\gamma }-1)}\frac{(1+e^{\gamma }\lambda z_{i})%
}{(1+e^{\gamma }\lambda z_{j})},  \label{Bethe eqs.}
\end{equation}%
which fix the spectrum of the parameters $z_{1},\ldots ,z_{P}$ (for details
see \cite{povolotsky mendes}). The set of  solutions $\mathcal{Z}$ of
(\ref{Bethe eqs.}) defines the sets of left and right eigenvectors
\begin{eqnarray}
\mathcal{B} &=&\left\{ \left\vert B_{Z}^{\gamma }\right\rangle ,Z\in
\mathcal{Z}\right\} ,  \label{eigenbases} \\
\overline{\mathcal{B}} &=&\left\{ \left\langle \overline{B}_{Z}^{\gamma
}\right\vert ,Z\in \mathcal{Z}\right\} .
\end{eqnarray}

A specific feature of the integrable models is that the same eigenvectors
diagonalize also a complete set of mutually commuting operators that are
as many as the degrees of freedom. The simplest example is the translation operator  $\mathbf{\tau }$ which translates a particle configuration one
step forward acting to the right%
\begin{equation}
\mathbf{\tau }\left\vert x_{1},\ldots ,x_{P}\right\rangle =\left\vert
x_{1}+1,\ldots ,x_{P}+1\right\rangle ,
\end{equation}%
while its adjoint action to the vector of the dual space is the one step
backward translation%
\begin{equation*}
\left\langle x_{1},\ldots ,x_{P}\right\vert \mathbf{\tau }=\left\langle
x_{1}-1,\ldots ,x_{P}-1\right\vert .
\end{equation*}%
Apparently the vectors from the sets $\mathcal{B}$ and $\overline{\mathcal{B}%
}$ are the eigenvectors of $\mathbf{\tau ,}$
\begin{eqnarray}
\mathbf{\tau }\left\vert B_{Z}^{\gamma }\right\rangle &=&\tau _{Z}\left\vert
B_{Z}^{\gamma }\right\rangle , \\
\left\langle \overline{B}_{Z}^{\gamma }\right\vert \mathbf{\tau } &=&\tau
_{Z}\left\langle \overline{B}_{Z}^{\gamma }\right\vert ,
\end{eqnarray}%
with the eigenvalue
\begin{equation}
\tau _{Z}=\left( z_{1}\cdots z_{P}\right) ^{-1}.
\end{equation}%
The translation by $L$ steps returns the system to itself, i.e.%
\begin{equation}
\mathbf{\tau }^{L}=\mathbf{E,}
\end{equation}%
where $\mathbf{E}$ is the identity operator. As a result $\tau _{Z}$ must be
an $L$-th root of unity.
\begin{equation}
\left( z_{1}\cdots z_{P}\right) ^{L}\equiv 1.  \label{translation invariance}
\end{equation}%
The same result can be obtained by multiplying all $L$ BE (\ref{Bethe eqs.}).

Assume that the sets $\mathcal{B}$ and $\overline{\mathcal{B}}$ are dual to each
other,
\begin{equation}
\left\langle \overline{B}_{Z}^{\gamma }|B_{Z^{\prime }}^{\gamma
}\right\rangle =\left\langle \overline{B}_{Z}^{\gamma }|B_{Z}^{\gamma
}\right\rangle \delta _{Z,Z^{\prime }},\,\,\,\,\left\langle \overline{B%
}_{Z}^{\gamma }|B_{Z}^{\gamma }\right\rangle \neq 0\,\,\,\,\,\,\,{\rm for\,\,any} \,\,Z,Z^{\prime }\in \mathcal{Z}
\end{equation}%
 and complete (i.e.
their cardinalities are as big as the dimension of the original space (\ref%
{dim(H_L)})). Then, the resolution of the identity relation holds%
\begin{equation}
\sum\limits_{Z\in \mathcal{Z}}\frac{\left\vert B_{Z}^{\gamma }\right\rangle
\left\langle \overline{B}_{Z}^{\gamma }\right\vert }{\left\langle \overline{B%
}_{Z}^{\gamma }|B_{Z}^{\gamma }\right\rangle }=\mathbf{E,}
\label{resolution of unity}
\end{equation}%
which in the configurational basis reads as follows%
\begin{equation}
\sum\limits_{Z\in \mathcal{Z}}\frac{\left\langle Y|B_{Z}^{\gamma
}\right\rangle \left\langle \overline{B}_{Z}^{\gamma }|X\right\rangle }{%
\left\langle \overline{B}_{Z}^{\gamma }|B_{Z}^{\gamma }\right\rangle }%
=\delta _{X,Y}.  \label{resolution of identity(basis)}
\end{equation}%
This allows us to derive the matrix element we are looking for.
\begin{eqnarray}
F_{t}^{\gamma }(X,t|X^{0},0) &=&\left\langle X|\mathbf{T}_{\gamma
}^{t}X^{0}\right\rangle =\sum\limits_{Z\in \mathcal{Z}}\frac{\left\langle X|%
\mathbf{T}_{\gamma }^{t}B_{Z}^{\gamma }\right\rangle \left\langle \overline{B%
}_{Z}^{\gamma }|X^{0}\right\rangle }{\left\langle \overline{B}_{Z}^{\gamma
}|B_{Z}^{\gamma }\right\rangle }  \nonumber \\
&=&\sum\limits_{Z\in \mathcal{Z}}\Lambda _{\gamma }^{t}(Z)\frac{\left\langle
X|B_{Z}^{\gamma }\right\rangle \left\langle \overline{B}_{Z}^{\gamma
}|X^{0}\right\rangle }{\left\langle \overline{B}_{Z}^{\gamma }|B_{Z}^{\gamma
}\right\rangle }.  \label{F_t}
\end{eqnarray}%
However the proof of completeness and orthogonality is a separate difficult
problem. An alternative way, which allows one to obtain the final formula (%
\ref{F_t}) without discussing these issues, is to prove the resolution of
the identity relation by a direct evaluation of the sum on the l.h.s. of (\ref%
{resolution of identity(basis)}). In our case, this problem turns out to be
solvable without explicit knowledge of the spectrum $\mathcal{Z}$%
.

\section{Location of the solutions of the Bethe equations in complex plane.%
\protect\bigskip}

Let us consider the system of BE (\ref{Bethe eqs.}). We introduce
for notational convenience  a new variable
\begin{equation}
z_{i}^{new}=z_{i}e^{\gamma }.  \label{y_i}
\end{equation}%
Below we will work only with these variables, so we omit
the superscript $"new"$ at $z_{i}$ avoiding an abuse
of notations.

Let us gather up into a single constant those parts of the equations (\ref{Bethe eqs.}), which have
no explicit dependence on the index $i$,
\begin{equation}
C=\left( -1\right) ^{P-1}e^{\gamma L}\prod\limits_{j=1}^{P}\frac{\left(
z_{j}-1\right) }{z_{j}\left( 1+z_{j}\lambda \right) },  \label{C}
\end{equation}%
Then the system of BE  (\ref{Bethe eqs.}) takes form of a unique polynomial
equation of degree $L$
\begin{equation}
z^{L-P}\left( z-1\right) ^{P}-C\left( 1+z\lambda \right) ^{P}=0.
\label{Bethe_C}
\end{equation}%
For specific values of $C$, those $P$ of $L$ roots of the polynomial, which
match the constraint (\ref{C}), give a solution of the Bethe equations.
The formal procedure of finding the solution was described in \cite{gwa
spohn}, \cite{golinelli mallick}. First, all $L$ roots of the polynomial
equation (\ref{Bethe_C}) ought to be found as functions of the parameter $C$%
. Then one substitutes any $P$ of them into the equation (\ref{C}),
obtaining a single equation. By solving this equation one obtains the solution
of the BE corresponding to a given set of the chosen roots.

Let us consider the analytic structure of the solutions in more detail. One
can rewrite (\ref{Bethe_C}) in the form
\begin{equation}
w(z)^{L}=C,  \label{w^L=C}
\end{equation}%
\ where the function $w(z)$ is
\begin{equation}
w(z)=\frac{z^{1-\rho }\left( z-1\right) ^{\rho }}{\left( 1+z\lambda \right)
^{\rho }}.
\end{equation}%
Equivalently one can write%
\begin{equation}
w(z)=\exp \left( 2\pi \mathrm{i} k/L\right) C^{1/L},  \label{w(z)=c^1/L}
\end{equation}%
where the integer $k$ is any integer chosen from the range $1\leq k\leq L$
specifying a particular choice of the branch. The branch of $C^{1/L}$ is
implied to be fixed, e.g.
\begin{equation*}
0\leq \arg (C^{1/L})<\frac{2\pi }{L}.
\end{equation*}

\begin{figure}[tbp]
\unitlength=1mm {\psfig{file=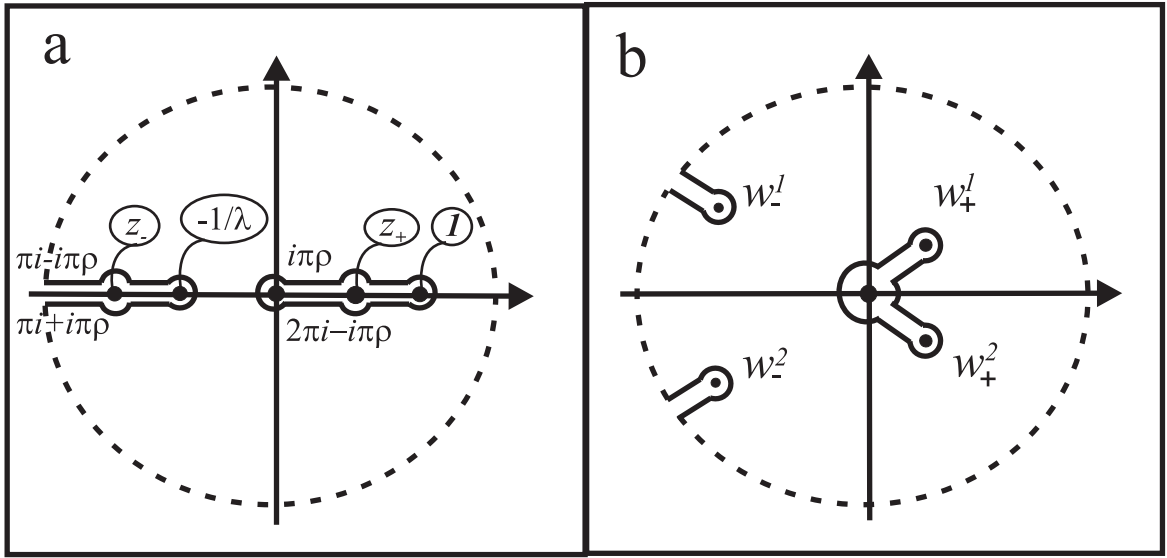}}
\caption{{}The domains $\mathbb{D}_{z}$ (a) and $\mathbb{D}_{w}$ (b).}
\label{complex plane}
\end{figure}

We use the notation $z_{+},z_{-}$ for the two solutions of the equation
\begin{equation}
\frac{\partial w\left( z\right) }{\partial z}=0,
\end{equation}%
which yields%
\begin{equation}
z_{\pm }=\frac{-1+\lambda -2\lambda \rho \pm \sqrt{(1+\lambda )(1+\lambda
(1-2\rho )^{2}))}}{2\lambda (1-\rho ).}.  \label{z+-}
\end{equation}%
We define the domain $\mathbb{D}_{z}$, (figure \ref{complex plane}a), as\ the
extended complex plane cut along the segments $[0,1]$ and $[-\infty
,-1/\lambda ]$ of the real axis and punctured at the points $%
\{0,1,-1/\lambda ,\infty ,z_{+},z_{-}\}$, and also the domain $\mathbb{D}%
_{w} $, ( figure \ref{complex plane}b), as the extended complex plane
punctured at the points $0,$ $\infty $, and at four points
\begin{eqnarray}
w_{+}^{1} &=&\left\vert w(z_{+})\right\vert e^{i\pi \rho }, \\
w_{+}^{2} &=&\left\vert w(z_{+})\right\vert e^{2\pi i-i\pi \rho }, \\
w_{-}^{1} &=&\left\vert w(z_{-})\right\vert e^{\pi i-i\pi \rho }, \\
w_{-}^{2} &=&\left\vert w(z_{-})\right\vert e^{\pi i+i\pi \rho },
\end{eqnarray}%
and cut along the straight segments: $\left[ 0,w_{+}^{1}\right] ,\left[
0,w_{+}^{2}\right] ,\left[ w_{-}^{1},\infty \right] ,\left[ w_{-}^{2},\infty %
\right] $. If we define $\mathbb{D}_{z}$ as a domain of $w(z)$, such
that the value of $\arg \left[ w(z)\right] $ is $\pi \rho $, $\pi \left(
2-\rho \right) $, and $\pi \left( 1-\rho \right) ,$ $\pi \left( 1+\rho
\right) $ at the upper and lower banks of the first and second branch cuts
of $\mathbb{D}_{z}$ respectively, then the mapping $w(z):$ $\mathbb{D}%
_{z}\rightarrow \mathbb{D}_{w}$ is the monovalued analytic mapping.
Going around the branch cut $[0,1]$ the value of $\arg \left[ w(z)\right] $
changes by $2\pi $ exactly once. Therefore, no repetition in the value of
$w(z)$ can occur, i.e. the mapping $w(z)$ is schlicht (one-sheet). Hence the
mapping $w(z)$ is bijective and one can construct an analytic, monovalued,
schlicht mapping $w^{-1}(\cdot ):\mathbb{D}_{w}\rightarrow \mathbb{D}_{z},$
inverse of$\ w(z)$. Then, given a complex number $a\in \mathbb{D}_{w}$,
the equation $w(z)=a$ has a unique simple root $z(a)\in \mathbb{D}_{z}$,
being an analytic function of $a$ in $\mathbb{D}_{w}$.

Thus, given a value of $C$, each root of (\ref{w^L=C}) is the unique simple
root of (\ref{w(z)=c^1/L}) for some $k$, and all the roots are different for
different $k$-s. To solve the Bethe equations one formally can choose $%
P$ integers $k_{1},\ldots ,k_{P}$ from the set $1,\ldots ,L$. Then,
substituting
\begin{equation}
z_{j}=w^{-1}\left[ \exp \left( 2\pi \mathrm{i}k_{j}/L\right) C^{1/L}\right]
\end{equation}%
into the constraint (\ref{C}) for all $j=1,\ldots ,P$, one obtains a unique
equation for the parameter $C$. Solving the equation for $C$, we obtain the
solutions corresponding to given set $k_{1},\ldots ,k_{P}$. Going through
all the possible sets of integers $k_{1},\ldots ,k_{P}$ we obtain all the
solutions of the Bethe equations.

It is easily seen that the eigenvectors (\ref{Bethe ansatz_right},\ref{bethe
ansatz left}) identically vanish on the solutions corresponding to the sets $%
k_{1},\ldots ,k_{P}$ containing a pair of two equal integers, $k_{i}=k_{j}$,
which result in $z_{i}=z_{j}$. Therefore one has to look over only those
sets $k_{1},\ldots ,k_{P}$, where all the numbers are different. If we
assume that for any set the equation for $C$ has exactly one solution
we obtain just as many solutions as we need (\ref{dim(H_L)}) to obtain the
complete set of linearly independent eigenvectors. The direct proof of this
fact however is beyond the aims of present article and will be considered
elsewhere.

\begin{remark}
For generic values of the parameter $\gamma $, and hence of the
parameter $C$, the r.h.s. of (\ref{w(z)=c^1/L}) is away from the branch
points of $w^{-1}(z)$, which ensures that the root of (\ref{w(z)=c^1/L}) is
simple. However, for specific values of $\gamma $ this can be not the case.
This does not create a problem as this situation can be considered
as the limiting case of the generic one. In particular, a double root can appear
at $z_{+}$ or $z_{-}$, as a consequence  of square root singularities
of $w^{-1}(z)$. In this case, one can think of it as a pair of
roots at different banks of the branch cut. This in fact specifies the way in which
it evolves when the value of $\gamma $ changes. Also the limit $C\rightarrow
0$, which implies $\gamma \rightarrow 0$, corresponds to $P$ roots meeting
at $z=1$. Then, the choice of different integers $k_{i}$ for different
roots $z_{i}$ removes the degeneracy as the arguments of $(z_{i}-1)$ are
different.
\end{remark}

Let us consider the curve $\Gamma _{c}$ defined by the equation
\begin{equation}
\left\vert w\left( z\right) \right\vert =c,  \label{cassini}
\end{equation}%
where $c$ is a real, positive number. All the roots are located on
curves where $c=\left\vert C^{1/L}\right\vert $, for some discrete set of
values of $C$ related to the solution via the equation (\ref{C}). The
particular case $\lambda =0$ of $\Gamma _{c}$ was named in \cite{golinelli
mallick} as the generalized Cassini oval. To describe the form of this curve, let
us first look at the behaviour of $\left\vert w\left( z\right) \right\vert $
at the real axis. It is easily seen that $\left\vert w\left( z\right)
\right\vert $ has two zeroes at the points $z=0$ and $z=1$, and diverges for
$z\rightarrow \pm \infty $ and $z\rightarrow -1/\lambda $. There are also
two extremums: $z_{+}$ and $\ z_{-}$ given by (\ref{z+-}), which are a minimum
and a maximum respectively. As $\rho $ varies from $0$ to $1$, the point $z_{+}$
monotonously moves from $1$ to $0$ and $z_{-}$ from $-1/\lambda $ to $%
-\infty $. At the same time $\left\vert w\left( z_{+}\right) \right\vert $
increases from zero for $0<\rho <1/2$, reaches the maximum at $\rho =1/2$ and
then decreases back to zero for $1/2<\rho <1$, while $\left\vert w\left(
z_{-}\right) \right\vert $ decreases from infinity for $0<\rho <1/2$, reaches the
minimum at $\rho =1/2$ and then increases back to infinity for $1/2<\rho <1$%
. Note that the values of $\left\vert w\left( z_{+}\right) \right\vert $ and
$\left\vert w\left( z_{-}\right) \right\vert $ meet only in one point $\rho
=1/2$ approaching its values from below and above respectively. Let us look
at the cross points of the plots $y=\left\vert w\left( x\right) \right\vert $
and $y=c$ as the constant $c$ grows starting from zero.
\begin{figure}[tbp]
\unitlength=1mm {
\psfig{file=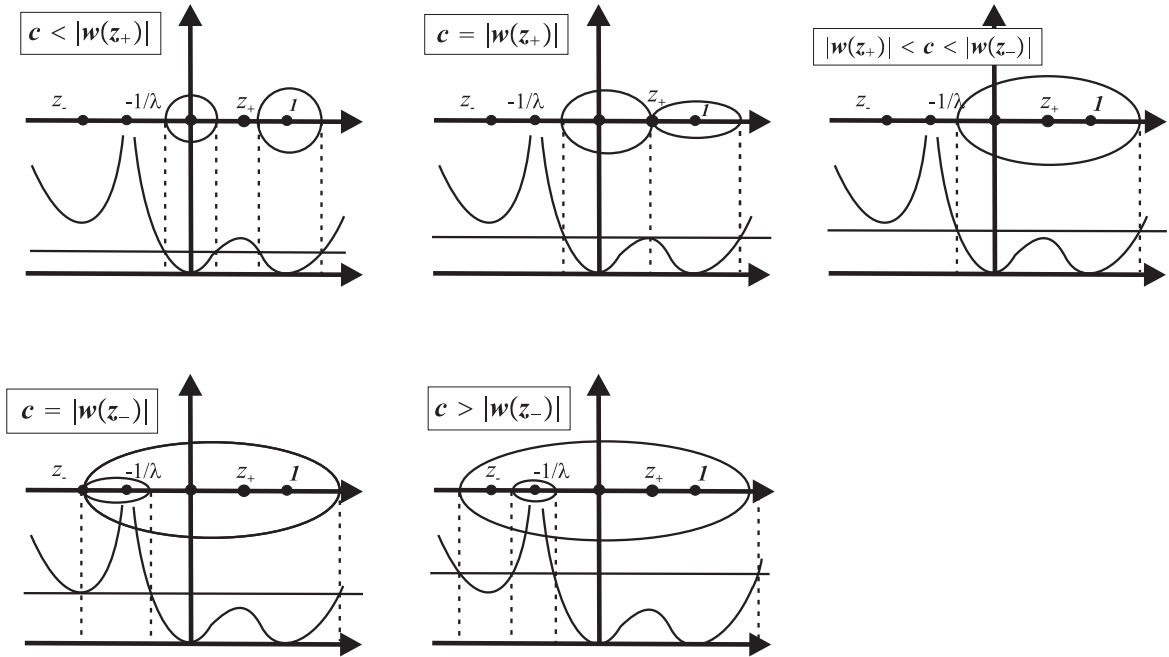,width=160mm}}
\caption{Schematic picture of the stages of the evolution of the contour $%
\Gamma _{c}$ as $c$ grows from zero to infinity. At the lower plots the
 the graphs $y=|w(x)|$ and $y=c$ are shown for each stage, which cross points
define the cross points of $\Gamma_c$ with the real axis. }
\label{stages}
\end{figure}
The following stages exist (see figure \ref{stages}):

\begin{description}
\item[$c<\left\vert w(z_{+})\right\vert $] For small $c$ there are four
cross points $\{ -1/\lambda <z_{1}<0,$ $0<z_{2}<z_{3}<1,$ $1<z_{4}\} $,
which corresponds to two "ovals" encircling the origin and the point $z=1.$
In the limit $c\rightarrow 0$ the form of the contours approaches circles of
vanishing radius collapsing to the points $z=0,1$. As the value of $c$
increases the point $z_{2}$ and $z_{3}$ move towards each other, and the
radius of the "ovals" increases .

\item[$c=\left\vert w(z_{+})\right\vert $] When the value of $c$ reaches $%
\left\vert w(z_{+})\right\vert $, the points $z_{2}$ and $z_{3}$ merge at $%
z_{+}$, i.e. the two "ovals" develop cusps meeting at $z_{+}$. Then the
shape of the curve resembles the lemniscate. This form survives, when
one considers the thermodynamic limit. Specifically, the right part of the
curve is the one considered in studies of low lying eigenstates \cite%
{povolotsky mendes}.

\item[$\left\vert w(z_{+})\right\vert <c<\left\vert w(z_{-})\right\vert $] %
As $c$ exceeds $\left\vert w(z_{+})\right\vert $ the crosspoints $z_{2}$,$%
z_{3}$ disappear after merging at $z_{+}$such that only the two, $%
\left\{ -1/\lambda <z_{1}<0,1<z_{4}\right\} $ remain. These are two
crosspoints with the horizontal axis of the "big oval", which appears after
the two "smaller ovals" merge. The big oval contains the points $z=0,1$
inside, while the point $z=-1/\lambda $ stays outside. As $c$ increases $%
z_{1},z_{4}$ move along the real axis towards $-1/\lambda $ and $+\infty $
respectively. As the point $z_{1}$ is confined between $0$ and $-1/\lambda $%
, while the effective radius of the "oval" is unbounded, it finally starts
to bend around $-1/\lambda $ and approaches the real axis from below and
above at the right of the point $-1/\lambda $.

\item[$c=\left\vert w(z_{-})\right\vert $] At this stage the two points at
the parts of the "oval" bending around $-1/\lambda $ meet at $z_{-}$, so
that two "ovals" appear, one inside the other. They have sharp cusps at
their only common point $z_{-}$ . The bigger "oval" goes around $-1/\lambda
,0,1$ while the smaller encircles only $-1/\lambda $.

\item[$c>\left\vert w(z_{-})\right\vert $] After forming two ovals at
previous stage, they detach from each other by splitting the common cross
point with the real axis at $z_{-}$ into two new cross points $z_{2}^{\prime
},z_{3}^{\prime }$, which then move along the real axis to $-\infty $ and $%
-1/\lambda $ respectively as $c$ go on growing. The form of the contours
then approach two circles of infinite and zero radia, the latter collapsing
to the point $-1/\lambda $.
\end{description}

In the case $\rho =1/2$, when the values of $\left\vert w\left( z_{+}\right)
\right\vert $ and $\left\vert w\left( z_{-}\right) \right\vert $ are equal,
the second and the fourth stages coincide while the third one
does not take place.

An important point of the above analysis is the limiting shape of the
curves under consideration as $c$ goes to zero or infinity. Specifically,
one can always choose a value of $c$ so small that the contour $\Gamma
_{c} $ separates a small neighborhood of the points $0$ and $1$
from the rest of the complex plane. For $c$ large, the same is valid for the
points $-1/\lambda $ and $\infty $. This fact will be used in the
next section to evaluate the sum over the Bethe roots.

\section{Summation over the Bethe roots.}

Let us rewrite the Bethe equations in terms of the variables $\left\{
z_{i}\right\} $, given by(\ref{y_i}), in polynomial form
\begin{equation}
P_{i}(Z)=0,  \label{polynomial BE}
\end{equation}%
where it is convenient to write the polynomials $P_{i}(Z)$ as a difference
of two other polynomials%
\begin{equation}
P_{i}(Z)\equiv g_{i}\left( Z\right) -h_{i}\left( Z\right)
\end{equation}%
which read as follows
\begin{eqnarray}
g_{i}\left( Z\right) &=&e^{-L\gamma }z_{i}^{L-P}\left( z_{i}-1\right)
^{P}\prod\limits_{j=1}^{P}z_{j}\left( 1+z_{j}\lambda \right) ,  \label{g(y)}
\\
h_{i}\left( Z\right) &=&\left( -1\right) ^{P-1}\left( 1+z_{i}\lambda \right)
^{P}\prod\limits_{j=1}^{P}\left( z_{j}-1\right) .  \label{h(y)}
\end{eqnarray}%
The aim of the present section is to evaluate the sum of the analytic
functions over the roots of the system (\ref{polynomial BE}). Two remarks
are necessary. First, we imply that the roots are bounded from infinity, as
otherwise the eigenvectors and eigenvalues would be singular. Second, the
polynomial equations can be satisfied with the solutions constructed from the
roots from the set $\{0,-1/\lambda ,1\}$. Furthermore, it is easy to see that
if one of the roots $z_{i}$ is taken from this set, the other $P-1$\ roots
must belong to this set as well. However, solutions constructed in this way
obviously do not match the constraint (\ref{translation invariance})
for $\gamma \neq 0$, and, therefore, must be excluded. Appearance of extra
solutions is due to multiplication of the BE by an expression that itself can
be zero, when transforming it to the polynomial form. In the case $\gamma
=0$ one such a solution exists. Namely it is $z_{1}=\cdots =z_{P}=1$, which
corresponds to the ground state of the evolution operator, i.e. the
stationary state of the stochastic process. As will be seen below, this case
can be treated as a limiting case of the generic situation and does not
require a special consideration.

Let us define the domain $D$ $\subset \mathbb{C}$ of the complex plane by
the inequalities
\begin{equation}
D=\left\{ z:a\leq \left\vert w(z)\right\vert \leq A\right\} ,  \label{D}
\end{equation}%
where $a$ and $A$ are two real positive constants, such that $a<A$. Denote
by $\Gamma _{c}$ the contour discussed in the previous section, defined by
(\ref{cassini})
\begin{equation}
\Gamma _{c}=\left\{ z:\left\vert w(z)\right\vert =c\right\} .
\end{equation}%
Then the domain $D$ is between $\Gamma_a$ and $\Gamma_A$, and its boundary%
\begin{equation}
\partial D=\left\{ z:\Gamma _{a}\cup \Gamma _{A}\right\}
\end{equation}%
is oriented in such a way that going along it in positive direction one
keeps the interior of $D$  left. Then we form a polycylinder domain $%
\mathbf{\Omega \subset }\mathbb{C}^{P}$ as a cartesian product of $P$ copies
of $D$
\begin{equation}
\mathbf{\Omega }=D_{1}\times \cdots \times D_{P}.
\end{equation}%
The skeleton $\mathbf{\Gamma }$ of \textbf{$\Omega $} is the subset of its
boundary $\partial \mathbf{\Omega }$ consisting of points, which are at
the boundary of every $D_{1},\ldots ,D_{P}$:
\begin{equation*}
\mathbf{\Gamma }=\partial D_{1}\times \cdots \times \partial D_{P}.
\end{equation*}%
The definition (\ref{D}) of $D$ guarantees that all the points $Z=\left(
z_{1},\ldots ,z_{P}\right) $, such that $z_{i}\in \{0,-1/\lambda ,1,\infty
\} $ for some $i$, are outside of \ \textbf{$\Omega $}. On\ the other hand,
the dimension of the complement of $\Omega $ approaches $P-1$ as $a$ and $A$
go to zero and infinity respectively, i.e. their Lebesgue measure in $%
\mathbb{C}^{P}$ vanishes. Thus, it is natural to expect that for $a$ small and $A$
large enough all the roots of the system (\ref{polynomial BE}) fall into $%
\mathbf{\Omega }$. Then, the sum of an analytic in $\mathbf{\Omega }$
function $f(Z)$ over the roots of BE which fall into $\mathbf{\Omega }$ can
be evaluated with the aid of the multi-dimensional logarithmic residue
theorem \cite{izenberg yuzhakov}.

\begin{theorem}
Let \textbf{$\Omega $}$\subset $ $\mathbb{C}^{P}$ be a polycylinder domain
with piecewise smooth boundary $\partial \mathbf{\Omega }$ and $\mathbf{%
\Gamma }$ be its skeleton. Let the mapping $\left\{ P_{i}(Z),i=1,\ldots
,P\right\} :\mathbf{\Omega }\rightarrow \mathbb{C}^{P}$ be holomorphic in $%
\mathbf{\Omega }$ and have no zeroes at the boundary $\partial \mathbf{%
\Omega }$. Then, for any function $f(Z)$ analytic in $%
\mathbf{\Omega }$, the sum of its values over the set $\mathcal{Z}=\left\{
Z\in \mathbf{\Omega :}P_{i}\left( Z\right) =0,i=1,\ldots ,P\right\} $ is
given by the following integral
\begin{eqnarray}
\sum_{Z\in \mathcal{Z}}f(Z) &=&\frac{1}{\left( 2\pi \mathrm{i}\right) ^{P}}%
\int_{\Gamma }f(Z)\frac{dP_{1}(Z)}{P_{1}(Z)}\wedge \cdots \wedge \frac{%
dP_{P}(Z)}{P_{P}(Z)}  \nonumber \\
&=&\int_{\Gamma }\frac{f(Z)}{\prod\nolimits_{i=1}^{P}P_{i}\left( Z\right) }%
\det \left[ \frac{\partial P_{i}\left( Z\right) }{\partial z_{j}}\right]
_{1\leq i,j\leq P}\prod\limits_{i=1}^{P}\frac{dz_{i}}{2\pi \mathrm{i}}.
\label{cauchy}
\end{eqnarray}%
Every zero is counted as many times as its multiplicity is.
\end{theorem}

The theorem is a particular case of the Caccippolly, Martinelly, Bishop,
Sorani theorem, see \cite{izenberg yuzhakov}. The original theorem is
proved for the domain called special analytic polyhedra, which in particular
ensures the absence of zeroes of the mapping at the boundary of domain,
which in turn guarantees that all zeroes inside the domain are the isolated
ones. In our case the absence of zeroes at the boundary $\partial \mathbf{%
\Omega }$ of the domain $\mathbf{\Omega }$, is provided by the following
lemma.

\begin{lemma}
Let $\mathbf{\Omega }$ be defined as above. Let $\lambda <1$. Then in the
range of $\gamma :\left( \rho ^{-1}\left\vert \ln \lambda \right\vert
>\gamma >0\right) $, there exist constants $a_{0}$ and $A_{0},$ such that
for any $a<a_{0}$ and $A>A_{0}$ the mapping $\left\{ P_{1}(Z),\ldots
,P_{P}(Z)\right\} $ has no zeroes on $\partial \mathbf{\Omega }$.
\end{lemma}

\begin{proof}
Suppose there is a point $Z=\left\{ z_{1},\ldots ,z_{P}\right\} \in \partial
\mathbf{\Omega }$, such that $P_{i}(Z)=0$ for $i=1,\ldots ,P$. It implies
that%
\begin{equation}
\left\vert \frac{g_{i}(Z)}{h_{i}\left( Z\right) }\right\vert \equiv
e^{-L\gamma }\left\vert w(z_{i})\right\vert ^{L}\left\vert \frac{z_{1}\cdots
z_{P}}{w(z_{1})\cdots w(z_{P})}\right\vert ^{1/\rho }=1.  \label{g(Y)/h(Y)}
\end{equation}%
The point $Z$ being at the boundary $\partial \mathbf{\Omega }$ of the
domain $\mathbf{\Omega }$ means that at least one coordinate $z_{i}$ from
the set $z_{1},\ldots ,z_{P}$ is at the boundary of $D_{i}$, i.e. is either
on $\Gamma _{a}$ or $\Gamma _{A}$. Let first
\begin{equation}
z_{i}\in \Gamma _{a}  \label{y_i in gamma_a}
\end{equation}%
for some $i$. As $z_{i}$ enters into (\ref{g(Y)/h(Y)}) only via $\left\vert
w(z_{i})\right\vert ^{L}$, while the other factors do not depend on the
index $i$ at all, $\left\vert w(z_{i})\right\vert ^{L}$ does not depend on $%
i $ either, i.e. (\ref{y_i in gamma_a}) holds for all $i=1,\ldots ,P,$ which
immediately yields
\begin{equation}
\left\vert z_{1}\cdots z_{P}\right\vert ^{1/\rho }=e^{L\gamma }.
\label{y1...yp}
\end{equation}
Recall that $a$ always can be chosen small enough such that $\Gamma _{a}$
belongs to small neighborhoods of the points $z=0$ and $z=1$. In
other words for any small $\varepsilon >0$ one can choose a small $a_{0}$ such
that for any $a<a_{0}$
\begin{equation}
\sup_{\left\{ z\in \Gamma _{a}\right\} }\left\vert z\right\vert
<1+\varepsilon .
\end{equation}%
Taking $\varepsilon <\left( e^{\gamma \rho }-1\right) $ we obtain
\begin{equation}
\left\vert z_{1}\cdots z_{P}\right\vert ^{1/\rho }<e^{\gamma L},
\end{equation}%
which contradicts (\ref{y1...yp}).

Consider now the case when $z_{i}\in \Gamma _{A}$ for some $i$. Analogously to
the previous case one needs to satisfy (\ref{y1...yp}) with the set of
solutions $z_{1},\ldots ,z_{P}$, which are on $\Gamma _{A}$, i.e. either go
to infinity or to $\left( -1/\lambda \right) $, as $A$ increases. Thus, for
any $\varepsilon >0$ one can always choose a large $A_{0}$ such that for any $%
A>A_{0} $%
\begin{equation}
\inf_{\left\{ z\in \Gamma _{A}\right\} }\left\vert z\right\vert >\left\vert
\frac{1}{\lambda }\right\vert -\varepsilon .
\end{equation}%
Taking $\varepsilon <\left( \left\vert 1/\lambda \right\vert -e^{\gamma \rho
}\right) $ we obtain
\begin{equation}
\left\vert z_{1}\cdots z_{P}\right\vert ^{1/\rho }>e^{\gamma L},
\end{equation}%
which also contradicts with \ref{y1...yp}.
\end{proof}

This lemma in particular ensures that the constants $a$ and $A$ always can be chosen small
and large respectively, so that all the solutions of BE, are inside $\mathbf{\Omega}$.
Indeed, according to the previous section the solutions always belong some contour
$\Gamma_c$ for a particular value of $c$.  Hence they would belong to the boundary of
$\mathbf{\Omega}$ defined with $c=a$ or $c=A$. As follows from the above lemma no
such solutions exist for $a<a_0$ and $A>A_0$. Therefore all the solutions are in
$\mathbf{\Omega}$ defined with such $a$ and $A$.

The next lemma provides a condition for evaluating the integral in (\ref%
{cauchy}) in the form of infinite series.
\begin{lemma}
\label{g/h><1}Let the domain $\mathbf{\Omega }$ and its skeleton $\Gamma $
be defined as above. Let the range of parameters be%
\begin{eqnarray}
0 &\leq &\lambda <1, \\
\rho ^{-1}\left\vert \ln \lambda \right\vert &<&\gamma <0
\end{eqnarray}%
and%
\begin{equation}
\rho \leq \frac{1}{2}.
\end{equation}%
Then, one can choose the constants $a$ and $A\ $such that for any $Z\in
\Gamma $ the following conditions hold:%
\begin{equation}
\left\vert \frac{g_{i}(Z)}{h_{i}\left( Z\right) }\right\vert <1
\label{g/h<1}
\end{equation}%
if $z_{i}\in \Gamma _{a}$ and%
\begin{equation}
\left\vert \frac{g_{i}(Z)}{h_{i}\left( Z\right) }\right\vert >1
\label{g/h>1}
\end{equation}%
if $z_{i}\in \Gamma _{A}$ Furthermore the limits $a\rightarrow 0$ and $%
A\rightarrow \infty $ can be taken simultaneously, in such a way, that these
inequalities hold.
\end{lemma}

\begin{proof}
It follows from the explicit form of $w(z)$ that for
any $\varepsilon >0$ one can choose $a$ so small that
\begin{eqnarray}
\inf_{z\in \Gamma _{a}}\left\vert z\right\vert &>&\left( 1-\varepsilon
\right) a^{\frac{1}{1-\rho }}  \label{sup y_a} \\
\sup_{z\in \Gamma _{a}}\left\vert z\right\vert &<&\left( 1+\varepsilon
\right) .
\end{eqnarray}%
On the other hand for any $\varepsilon >0$ one
can choose $A$ so large that
\begin{eqnarray}
\inf_{z\in \Gamma _{A}}\left\vert z\right\vert &>&\frac{(1-\varepsilon )}{%
\lambda }  \label{sup y_A} \\
\sup_{z\in \Gamma _{A}}\left\vert z\right\vert &<&\left( 1+\varepsilon
\right) \lambda ^{\frac{\rho }{1-\rho }}A^{\frac{1}{1-\rho }} .
\end{eqnarray}%
Let us write $c_{i}=a$ if $z_{i}\in \Gamma _{a}$ and $c_{i}=A$ if $z_{i}\in
\Gamma _{A}$. Then we have
\begin{equation}
\left\vert \frac{g_{i}(Z)}{h_{i}\left( Z\right) }\right\vert =e^{-L\gamma
}c_{i}^{L}\left\vert \frac{z_{1}\cdots z_{P}}{c_{1}\cdots c_{P}}\right\vert
^{1/\rho },
\end{equation}%
Now we can use the inequalities (\ref{sup y_a}) and (\ref{sup y_A}) to
estimate the bounds for the ratios $\left\vert z_{i}/c_{i}\right\vert $,
which yield
\begin{equation}
\hspace{-20mm}~\left[ \left( 1-\varepsilon \right) \min \left( a^{\frac{\rho }{1-\rho }},%
\frac{1}{\lambda A}\right) \right] ^{P}<\left\vert \frac{z_{1}\cdots z_{P}}{%
c_{1}\cdots c_{P}}\right\vert ^{1/\rho }<\left[ \left( 1+\varepsilon \right)
\max \left( \left( \lambda A\right) ^{\frac{\rho }{1-\rho }},\frac{1}{a}%
\right) \right] ^{P}.
\end{equation}%
If we choose $A$ and $a$ such that%
\begin{equation}
\left( \lambda A\right) ^{\frac{\rho }{1-\rho }}\leq \frac{1}{a},
\label{A<1/a}
\end{equation}%
then for $c_{i}=a$ we have
\begin{equation}
e^{-L\gamma }c_{i}^{L}\left\vert \frac{z_{1}\cdots z_{P}}{c_{1}\cdots c_{P}}%
\right\vert ^{1/\rho }<\left( 1+\varepsilon \right) ^{L}e^{-L\gamma }<1,
\label{conv1}
\end{equation}%
which is (\ref{g/h<1}). At the same time if we choose them such that
\begin{equation}
a^{\frac{\rho }{1-\rho }}\geq \frac{1}{\lambda A},  \label{A>1/a}
\end{equation}%
then for $c_{i}=A$ we have
\begin{equation}
e^{-L\gamma }c_{i}^{L}\left\vert \frac{z_{1}\cdots z_{P}}{c_{1}\cdots c_{P}}%
\right\vert ^{1/\rho }>e^{-L\gamma }\left( \frac{1-\varepsilon }{\lambda }%
\right) ^{L}>1,  \label{conv2}
\end{equation}%
which is (\ref{g/h>1}). The only what we need is to satisfy both conditions (%
\ref{A<1/a}) and (\ref{A>1/a}) simultaneously, which implies%
\begin{equation}
\left( \lambda A\right) ^{-1}\leq a^{\frac{\rho }{1-\rho }}\leq \left(
\lambda A\right) ^{-\left( \frac{\rho }{1-\rho }\right) ^{2}}.
\end{equation}%
This is possible if $\rho \leq 1/2$.
\end{proof}

\begin{remark}
The limit $\gamma \rightarrow 0$ in two above lemmas can be considered after
the limits $a\rightarrow 0$ and $A\rightarrow \infty $ are taken. This
particularly solves the problem of the groundstate mentioned above. While
for $\gamma >0$ the roots corresponding to the groundstate are away from $%
Z=(1,\ldots ,1)$, they approach this point when $\gamma $ approaches zero.
However due to the order of limits described they still remain separated
from this point by the boundary of $\mathbf{\Omega }$.
\end{remark}

Using lemma \ref{g/h><1}, we can represent the factor $1/P_{i}(Z)$
under the integral in (\ref{cauchy}) in the form of infinite sum. Indeed, for $%
z_{i}\in \Gamma _{a}$ we have
\begin{equation}
\frac{1}{P_{i}(Z)}=\frac{1}{g_{i}\left( Z\right) -h_{i}\left( Z\right) }=%
-\frac{1}{h_{i}\left( Z\right) }\sum\limits_{n=0}^{\infty }\left( \frac{%
g_{i}\left( Z\right) }{h_{i}\left( Z\right) }\right) ^{n},
\label{1/P_i  Gamma_a}
\end{equation}%
while for $z_{i}\in \Gamma _{A}$
\begin{equation}
\frac{1}{P_{i}(Z)}=\frac{1}{g_{i}\left( Z\right) -h_{i}\left( Z\right) }=%
\frac{1}{h_{i}\left( Z\right) }\sum\limits_{n=-\infty }^{-1}\left( \frac{%
g_{i}\left( Z\right) }{h_{i}\left( Z\right) }\right) ^{n}.
\label{1/P_i Gamma_A}
\end{equation}%
Thanks to (\ref{conv1},\ref{conv2}), the series are absolutely and uniformly
convergent, and, as such, can be integrated term by term. Note that the
summands have no singularities in $\mathbf{\Omega }$. Therefore the
contours $\Gamma _{a}$ and $\Gamma _{A}$ can be deformed into a single
contour. With respect to singularities of the expression under the integral,
it has the same form as the contour $\Gamma ^{\infty }$ described in the
Section~\ref{Sec Bethe ansatz}, which was used to construct the resolution of the identity
operator in the case of infinite lattice. Thus, both the contours $\Gamma _{a}$
and $\Gamma _{A}$ can be deformed into $\Gamma ^{\infty }$ and the term
corresponding to the integral over $\Gamma _{a}$ brings a minus sign due
to the opposite orientation. The final expression has no dependence on
the values of $a$ and $A$. Therefore, assuming the limit $a\rightarrow
0,A\rightarrow \infty $ one can think of $\Omega $ as of product of $P$
complex plains punctured in four points $\left\{ 0,1,-1/\lambda ,\infty
\right\} $, which clearly contains all the necessary solutions of the BE. As
a result we have the following expression for the sum over the roots of the
Bethe equations.

\begin{theorem}
The sum of the values of a function $f(Z)$ analytic in the whole complex
plane, except maybe the points $\left\{ 0,1,-1/\lambda ,\infty \right\}$,
over the roots of the Bethe equations is given by the following $P$-tuple
absolutely converging sum.
\begin{eqnarray}
\sum_{Z\in \mathcal{Z}}f(Z) &=&\sum\limits_{n_{1}=-\infty }^{\infty }\cdots
\sum\limits_{n_{P}=-\infty }^{\infty }\int_{\Gamma _{1}^{\infty }\times
\cdots \times \Gamma _{P}^{\infty }}f(Z)\prod\nolimits_{i=1}^{P}\left( \frac{%
h_{i}(Z)}{g_{i}(Z)}\right) ^{n_{i}}  \nonumber \\
&&\times \det \left[ \frac{\partial \left( \ln g_{k}(Z)-\ln h_{k}(Z)\right)
}{\partial z_{j}}\right] _{1\leq k,j\leq P}\prod\limits_{l=1}^{P}\frac{dz_{l}%
}{2\pi \mathrm{i}},  \label{series representation}
\end{eqnarray}
\end{theorem}

Note that we write $\partial \left( \ln g_{k}(Z)\right)
/\partial z_{j}\equiv \left( \partial g_{k}(Z)/\partial
z_{j}\right) /g_{k}(Z)$ in the determinant instead of $\left( \partial g_{k}(Z)/\partial z_{j}\right)
/h_{k}(Z)$, because the equality $h_{k}(Z)=g_{k}(Z)$ holds
on the roots of BE, which are the only contributing the
integral.

\section{Proof of the resolution of the identity  and  formula for the
transition probability}

Now, we are in a position to write the sum in (\ref{resolution of
identity(basis)}) in the integral form (\ref{series representation})
substituting%
\begin{equation}
f(Z)=\frac{\left\langle X|B_{Ze^{-\gamma }}^{\gamma }\right\rangle
\left\langle \overline{B}_{Ze^{-\gamma }}^{\gamma }|Y\right\rangle }{%
\left\langle \overline{B}_{Ze^{-\gamma }}^{\gamma }|B_{Ze^{-\gamma
}}^{\gamma }\right\rangle }.  \label{f(z)}
\end{equation}%
To this end we need the expression for the norm $\left\langle \overline{B}%
_{Ze^{-\gamma }}^{\gamma }|B_{Ze^{-\gamma }}^{\gamma }\right\rangle $. The
hypothesis about the form of the norms of Bethe vectors was first proposed
by Gaudin \cite{Gaudin,Gaudin M McCoy B M and Wu T T}. Later it was proved
by Korepin \cite{Korepin} within the quantum
inverse scattering method for XXX and XXZ type models. Those results can be
applied to our model with minor changes. However, they require developing
the quantum inverse scattering method, which is not a subject of the present
article, and will be done elsewhere. Here we use the formulas as they are
given in the Korepin's article. Note that the proof below
and of the final result do not rely on the validity of the formula
for the norm. The latter serves as a hint for writing the expression under
the integral, while the validity of the results follows from the resolution
of the identity proof made independently. Written in the the transformed
variables (\ref{y_i}), Gaudin formula yields
\begin{equation}
\left\langle \overline{B}_{Ze^{-\gamma }}^{\gamma }|B_{Ze^{-\gamma
}}^{\gamma }\right\rangle =\det \left[ z_{i}\frac{\partial }{\partial z_{i}}%
\ln \frac{g_{j}(Z)}{h_{j}(Z)}\right] _{i,j=1,\ldots ,P},
\end{equation}%
which being in the denominator of the expression under the integral cancels
the Jacobian in the numerator, yielding only the factor $\left( z_{1}\cdots
z_{P}\right) ^{-1}$. Substituting the explicit form of the eigenvectors (\ref%
{Bethe ansatz_right},\ref{bethe ansatz left}) and the functions $g(z)$ and $%
h(z)$, (\ref{g(y)},\ref{h(y)}), and performing one summation over the
permutations, which is trivial due to the permutation symmetry of the
summands, we come to the following expression.
\begin{eqnarray}
&\hspace{-25mm}~\sum\limits_{Z\in \mathcal{Z}}\frac{\left\langle X|B_{Z}^{\gamma
}\right\rangle \left\langle \overline{B}_{Z}^{\gamma }|Y\right\rangle }{%
\left\langle \overline{B}_{Z}^{\gamma }|B_{Z}^{\gamma }\right\rangle }
=W(X)\sum\limits_{n_{1}=-\infty }^{\infty }\cdots
\sum\limits_{n_{P}=-\infty }^{\infty }\left( -1\right)
^{(P-1)\sum_{k=1}^{P}n_{k}}\sum\limits_{\sigma \in S_{P}}\left( -1\right)
^{|\sigma |}  \label{resolution_integral} \\
&\times \prod\nolimits_{i=1}^{P}\int_{\Gamma ^{\infty }}\left( \frac{%
1+\lambda z_i}{1-1/z_i}\right) ^{i-\sigma _{i}+Pn_{i}-\sum_{l=1}^{P}n_{l}}\left(
e^{-\gamma }z_i\right) ^{-x_{i}+y_{\sigma_i}-Ln_{i}}\frac{dz_i}{2\pi iz_i}.  \nonumber
\end{eqnarray}%
Though there is an infinite $P$-tuple sum, it turns out that only few terms in each sum
contribute. The following lemma establishes which summands are nonzero.

\begin{lemma}
\label{nonzero terms}Let $X,Y\in \mathbb{Z}_{<,L}^{P}$ be two particle
configurations, $\sigma $ be a permutation $\left( \sigma _{1},\ldots
,\sigma _{P}\right) $ of the integers $1,\ldots ,P$, $\left\{ n_{1},\ldots
,n_{P}\right\} \in \mathbb{Z}^{P}$ be a set of integers. Then, the necessary
conditions for the product
\begin{equation}
\prod\nolimits_{i=1}^{P}\int_{\Gamma ^{\infty }}\left( \frac{1+\lambda z_i}{%
1-1/z_i}\right) ^{i-\sigma _{i}+Pn_{i}-\sum_{k=1}^{P}n_{k}}\left( e^{-\gamma
}z_i\right) ^{-x_{i}+y_{\sigma _{i}}-Ln_{i}}\frac{dz_i}{2 \pi \mathrm{i} z_i}
\end{equation}%
to be nonzero are
\begin{equation}
n_{i}\in \left\{ -1,0,1\right\}
\end{equation}%
for $i=1,\ldots ,P$. \ Furthermore, the cases when $n_{i}=\pm 1$ for some $%
i=1,\ldots ,P,$ suggest that
\begin{equation*}
\sum_{i=1}^{P}n_{i}=0
\end{equation*}%
and there are clusters in the configurations $X$ and $Y$, which in
particular contain the sites $1$ and $L$.
\end{lemma}

\begin{proof}
To evaluate the integrals over $\Gamma ^{\infty }$ under the product we
expand the expression under the integral into the Laurent series in the ring
$1<\left\vert z\right\vert <1/\lambda $, implying $\lambda >1$, and look for
the coefficient coming with $z^{-1}$. Let us for brevity introduce the
notations%
\begin{eqnarray}
a_{ij} &=&i-j+Pn_{i}-\sum_{k=1}^{P}n_{k},  \label{a_ij} \\
b_{ij} &=&x_{i}-y_{j}+Ln_{i}.  \label{b_ij}
\end{eqnarray}%
Then, the integrals to be nonzero, the following conditions must be met:%
\begin{eqnarray}
\textrm{a. \ \ \ if \ }a_{i\sigma _{i}} &=&0,\textrm{ \ then \ }b_{i\sigma
_{i}}=0;  \label{a} \\
\textrm{b. \ \ \ if \ }a_{i\sigma _{i}} &>&0,\textrm{ \ then \ }b_{i\sigma
_{i}}\leq a_{i\sigma _{i}};  \label{b} \\
\textrm{c. \ \ \ if \ }a_{i\sigma _{i}} &<&0,\textrm{ \ then \ }b_{i\sigma
_{i}}\geq a_{i\sigma _{i}}.  \label{c}
\end{eqnarray}%
Given particle configurations $X,Y$ and a permutation $\sigma $, every
element from the set $\{n_{i}\}_{i=1,\ldots ,P}$ fall into one of three
classes (a),(b) or (c) depending on the sign of $a_{i\sigma _{i}}$. Then the
inequalities (\ref{a}-\ref{c}) determine which restrictions on $n_{i}$ must
be imposed, all the integrals in the product to be nonzero simultaneously.

First we note that as the coordinates of particles on the finite lattice
satisfy $\left\vert x_{i}-y_{j}\right\vert <L$, the only way to satisfy the
equality $b_{i\sigma _{i}}=0$ is to put $n_{i}=0$. Therefore, this is always
the case for the numbers $n_{i}$, which belong to the class (a).

One can use a similar argument to consider a particular case of the set $%
\{n_{i}\}_{i=1,\ldots ,P}$ with all components equal.
\begin{equation}
n_{1}=n_{2}=\ldots =n_{P}\equiv n.
\end{equation}%
Then for such $i$ that $a_{i\sigma _{i}}=\left( i-\sigma _{i}\right) >0$,
class (b), (\ref{b}) requires $n\leq 0$, while for such $j$ that $a_{j\sigma
_{j}}=\left( j-\sigma _{j}\right) <0$, class (c), (\ref{c}) implies $n\geq 0$%
. Apparently for any permutation $\sigma $, where for some $i$ there is a
number $\sigma _{i}$ which satisfies $\left( i-\sigma _{i}\right) \geq 0$,
there must be at least one $j$ such that $\left( j-\sigma _{j}\right) \leq 0$%
. Thus we necessarily have $n=0$.

To study the other sets we estimate the bounds of their maximal and minimal
elements. Consider the sets $\{n_{i}\}_{i=1,\ldots ,P}$, where not all
components are equal. From such set a maximal element can be chosen
\begin{equation}
n_{\max }\equiv \max \{n_{1},\ldots ,n_{P}\}.
\end{equation}%
Of course, in general several numbers from the set can attend the maximum.
The proof below consists of two steps. We first show that at least one of
them falls either into the class (a) or into the class (b), i.e. satisfies
either the first equality in (\ref{a}) or the first inequality in (\ref{b}).
Then we find which restrictions on $n_{\max}$follow from the second
equality of (\ref{a}) or the second inequality (\ref{b}).

To proceed with the first step we note that for any $k$ such that $n_{k}=$ $%
n_{\max }$ at least a part of $a_{k\sigma _{k}}$ is always positive%
\begin{equation}
Pn_{k}-\sum_{l=1}^{P}n_{l}>0.
\end{equation}%
The remaining term $\left( k-\sigma _{k}\right) $, which also enters into $%
a_{k\sigma _{k}}$, is either positive or negative. In the case when it is
positive or it is negative, but its absolute value is smaller then one
of the other terms, we have $a_{k\sigma _{k}}\geq 0$. It can happen, however,
that $\left( k-\sigma _{k}\right) $ is negative and its absolute value is
large enough to turn $a_{k\sigma _{k}}$ to be negative. This is possible
when
\begin{equation}
Pn_{k}-\sum_{k=1}^{P}n_{k}\leq \left\vert k-\sigma _{k}\right\vert \leq P-1.
\end{equation}%
Let $s$ be an integer, $1\leq s\leq P$, such that
\begin{equation}
Pn_{\max }-\sum_{k=1}^{P}n_{k}=P-s.
\end{equation}%
The last equation suggests that the number elements in the set $%
\{n_{i}\}_{i=1,\ldots ,P}$ equal to $n_{\max }$ is at least $s$, i.e. there
exists at least $s$ integers $1<k_{1},<\cdots <k_{s},\leq P$ such that
\begin{equation}
n_{k_{1}}=\cdots =n_{k_{s}}=n_{\max}.
\end{equation}%
Let us now check the sign of $a_{i\sigma _{i}}$ for $i$ chosen from $%
k_{1},\ldots ,k_{s}$ and an arbitrary permutation $\sigma $. At worst we
have $\sigma _{k_{1}}=P-s+1,\ldots ,\sigma _{k_{s}}=P$, in which case for $%
i=k_{1}$, $a_{i\sigma _{i}}=\left[ k_{1}-\left( P-s+1\right) \right] +\left(
P-s\right) \geq 0$. \ For any other $\sigma $ either $\sigma _{i}$ can be
taken smaller or $i$ bigger, so the inequality becomes strict. Thus,
we conclude that there is always at least one maximal element belonging
either to the class (a) or (b), which finishes the first step of the proof.

According to the above arguments for the class (a) we get $n_{\max }=0$. Let
now $n_{i}=n_{\max }$ be in the class (b). Combination of
the second inequality (\ref{b}) and the finite lattice restriction on
the difference of the particle coordinates
\begin{equation}
\left\vert \left( x_{i}-y_{j}\right) -\left( i-j\right) \right\vert \leq L-P
\label{finit lattice.}
\end{equation}%
give the upper bound for $n_{\max }$.
\begin{equation}
n_{\max }\leq 1-\frac{1}{L-P}\sum_{k=1}^{P}n_{k}.  \label{n_max<=}
\end{equation}%
All the above arguments can be applied also to the minimal element as well%
\begin{equation*}
n_{\min }\equiv \min \{n_{1},\ldots ,n_{P}\},
\end{equation*}%
so we conclude that it belongs either to the class (a), where
$n_{\min }=0$ or to the class (c), where (\ref{c}) and (\ref{finit
lattice.}) yield
\begin{equation}
n_{\min }\geq -1-\frac{1}{L-P}\sum_{k=1}^{P}n_{k}.  \label{n_min>=}
\end{equation}%
It follows from (\ref{n_max<=}) that if $\sum_{k=1}^{P}n_{k}$ is positive,
then $n_{\max }$ must be strictly less than $1$ and hence nonsensitive. At
the same time, one sees  from (\ref{n_max<=}) that if $\sum_{k=1}^{P}n_{k}$ is
negative then $n_{\min }$ must be strictly greater than $-1$ and hence
nonnegative. Thus the only possibility is to have
\begin{equation}
\sum_{k=1}^{P}n_{k}=0.  \label{sum n=0}
\end{equation}%
Substituting this back into (\ref{n_max<=},\ref{n_min>=}) we conclude that
there are only three possible values of the elements of the set $%
\{n_{i}\}_{i=1,\ldots ,P}$
\begin{equation}
n_{i}\in \{-1,0,1\}, \, \, i=1,\ldots ,P,
\end{equation}%
the numbers $+1$ and $-1$ appearing in the set $\{n_{i}\}_{i=1,\ldots ,P}$
equally many times.

Consider now the case when not all $n_{i}$ are equal to zero, i.e.
\begin{equation}
n_{\min }=-1,n_{\max }=1  \label{1}
\end{equation}%
If we return to the inequalities (\ref{b},\ref{c}) and repeat the derivation
of (\ref{n_max<=},\ref{n_min>=}) using (\ref{sum n=0}), we obtain that the
case (\ref{1}) is realized when the weak inequality (\ref{finit lattice.})
turns to the equality. Specifically, deriving the estimate for $%
n_{i}=n_{\max }=1$, the necessary condition for $n_{\max }$ to be equal to $%
1 $ is
\begin{equation}
y_{\sigma _{i}}-x_{i}-(\sigma _{i}-i)=L-P.
\end{equation}%
This means that the particles with coordinates $x_{i}$ and $y_{\sigma _{i}}$
belong to the clusters in $X$ and $Y$, which spread to the last and the
first sites respectively, i.e. the particle at $x_{i}$ belongs to the
cluster of $X$, which starts with $x_{1}=1$ and the particle at $y_{\sigma
_{i}}$ belongs to the cluster of $Y$, which ends with $y_{L}=L$. Similarly,
for $n_{j}=n_{\min }=-1$ we have
\begin{equation*}
y_{\sigma _{j}}-x_{j}-(\sigma _{j}-j)=-\left( L-P\right) ,
\end{equation*}%
i.e. the particle at $x_{j}$ belongs to the cluster of $X$, which ends with $%
x_{L}=L$ and the particle at $y_{\sigma _{j}}$belongs to the cluster of $Y$,
which starts with $y_{1}=1$. Thus we conclude that in both $X$ and $Y$ there
exist the clusters, which cover at least the sites at the positions $1$ and $%
L $. This proves the last statement of the lemma.
\end{proof}

Of course the choice of the reference point at the ring has a conventional
character. One can get reed of the terms $n_{i}=\pm 1$ by simple
rotation, which places any hole of one of the configurations $X$ or $Y$
either into the site $1$ or $L$. This fact is useful for the proof of the
resolution of the identity relation.

\begin{theorem}
The resolution of the identity operator is given by l.h.s of (\ref%
{resolution of unity}).
\end{theorem}

\begin{proof}
Consider the integral representation (\ref{resolution_integral}) of the sum
in (\ref{resolution of unity}). According to the pervious lemma if one of
the configurations $X$ and $Y$ are such that any of the sites $1$ and $L$ is
empty the only term of the sum that remains corresponds to $n_{1}=\cdots
=n_{P}=0$. This term coincides with the resolution of the identity operator
for the infinite lattice. For the proof of the infinite lattice case we
refer the reader to the Proposition 1 from our first paper \cite{povolotsky
priezzhev}.

The cases where in both configurations $X$ and $Y$ there is a cluster
containing the sites $1$ and $L$ can be reduced to the previous situation by
translation. Specifically we note that the product $\left\langle
X|B_{Z}^{\gamma }\right\rangle \left\langle \overline{B}_{Z}^{\gamma
}|Y\right\rangle $ is invariant under the translations, i.e.
\begin{eqnarray*}
&&\left\langle x_{1}+1,\ldots ,x_{P}+1|B_{Z}^{\gamma }\right\rangle
\left\langle \overline{B}_{Z}^{\gamma }|y_{1}+1,\ldots ,y_{P}+1\right\rangle
\\
&=&\left\langle X\mathbf{\tau }^{-1}|B_{Z}^{\gamma }\right\rangle
\left\langle \overline{B}_{Z}^{\gamma }|\mathbf{\tau }Y\right\rangle =\tau
_{Z}\tau _{Z}^{-1}\left\langle X|B_{Z}^{\gamma }\right\rangle \left\langle
\overline{B}_{Z}^{\gamma }|Y\right\rangle \\
&=&\left\langle x_{1},\ldots ,x_{P}|B_{Z}^{\gamma }\right\rangle
\left\langle \overline{B}_{Z}^{\gamma }|y_{1},\ldots ,y_{P}\right\rangle .
\end{eqnarray*}%
Hence we can repeatedly apply the translation operators $\mathbf{\tau }$ and
$\mathbf{\tau }^{-1}$ until a hole comes either to the site $1$ or $L$. In
this way the problem is reduced to the proved one.
\end{proof}

Since we have proven the formula for the resolution of the identity
operators, we can apply it to find the matrix element we are looking for. To
this end, as shown in (\ref{F_t}) we must insert the eigenvalue of $\mathbf{T%
}_{\gamma }^{t}$ under the integral. Then the integral representation of
$F_{t}^{\gamma }(X,t|Y,0)$ is
\begin{eqnarray}
\left\langle X|\mathbf{T}_{\gamma }^{t}Y\right\rangle &=&\left( 1+\lambda
\right) ^{-Pt}W(X)\sum\limits_{\sigma \in S_{P}}\left( -1\right) ^{|\sigma |}
\nonumber \\
&&\times \int_{\Gamma }\prod\nolimits_{i=1}^{P}\frac{\left( e^{-\gamma
}z_{i}\right) ^{-x_{i}+y_{\sigma _{i}}}\left( 1+\lambda z_{i}\right)
^{i-\sigma _{i}+t}}{\left( 1-1/z_{i}\right) ^{i-\sigma _{i}}}\frac{%
h_{i}\left( Z\right) }{P_{i}\left( Z\right) }\frac{dz_{i}}{2\pi \mathrm{i}z_{i}}.
\end{eqnarray}%
Going to the form of the $P$-tuple sum we have
\begin{eqnarray}
\hspace{-15mm}\left\langle X|\mathbf{T}_{\gamma }^{t}Y\right\rangle
=W(X)\sum\limits_{n_{1}=-\infty }^{\infty }\cdots
\sum\limits_{n_{P}=-\infty }^{\infty }e^{\gamma \sum_{j=1}^{P}\left(
x_{j}-y_{j}+Ln_{j}\right) }\left( -1\right)
^{(P-1)\sum_{l=1}^{P}n_{l}}  \nonumber \\
\times\sum\limits_{\sigma \in S_{P}}\left( -1\right)
^{|\sigma |} \prod\nolimits_{i=1}^{P}f(i-\sigma
_{i}+Pn_{i}-\sum_{k=1}^{P}n_{k},x_{i}-y_{\sigma_i}+Ln_{i},t),  \label{<T^t_gamma>}
\end{eqnarray}%
where%
\begin{equation}
f(a,b,t)=\left( 1+\lambda \right) ^{-t}\int_{\Gamma ^{\infty }}\left(
1+\lambda z\right) ^{t}\left( \frac{1+\lambda z}{1-1/z}\right) ^{a}z^{-b}%
\frac{dz}{2\pi \mathrm{i}z}.
\end{equation}%
The integral for $f(a,b,t)$ is evaluated in terms of the hypergeometric
functions (\ref{f(a,b,t)}). The sum over the permutations leads us to the
determinant, given in (\ref{F(X,Y,t)_ring}). By putting $\gamma =0$ we
obtain the result for conditional probability announced.

Like the sum obtained for the resolution of the identity operator, which is
the particular case of the sum (\ref{<T^t_gamma>}) at $t=0$, the latter sum
being formally infinite, however contains finitely many nonzero terms. The
analysis similar to one of the lemma \ref{nonzero terms} shows that at
time $t$ the upper bound for the maximal element $n_{\max }(t)$ of the set $%
\{n_{i}\}_{i=1,\ldots ,P}$, which ensures corresponding summand to be
nonzero, is
\begin{equation}
n_{\max }(t)\leq 1-\frac{1}{\left( L-P\right) }\sum_{k=1}^{P}n_{k}+\frac{t}{%
\left( L-P\right) },  \label{nmax(t)<=}
\end{equation}%
while that for minimal one, $n_{\min }(t)$, is still like it was for the
$t=0$ case
\begin{equation}
n_{\min }\left( t\right) \geq -1-\frac{1}{\left( L-P\right) }%
\sum_{k=1}^{P}n_{k}.  \label{nmin(t)>=}
\end{equation}%
The bounds for the sum $\sum_{k=1}^{P}n_{k}$ can be obtained from the
following arguments. Suppose that $\sum_{k=1}^{P}n_{k}>t$. Then (\ref%
{nmax(t)<=}) requires $n_{\max }(t)\leq 0$, which contradicts the
assumption. Thus we have
\begin{equation}
\sum_{k=1}^{P}n_{k}\leq t.
\end{equation}%
Another argument can be given, based on the fact that $\sum_{k=1}^{P}n_{k}%
\leq Pn_{\max }(t)$. Then using (\ref{nmax(t)<=}) we obtain%
\begin{equation}
\sum_{k=1}^{P}n_{k}\leq P+\frac{t-P}{L}.
\end{equation}%
The first upper bound is lower than the second, when $t<P,$ and vice versa
when $t>P$. For the lower bound we suppose that $\sum_{k=1}^{P}n_{k}<0$,
which contradicts (\ref{nmin(t)>=}) and results in%
\begin{equation}
\sum_{k=1}^{P}n_{k}\geq 0.
\end{equation}%
Finally we can use these inequalities to estimate the range of the
summation indices corresponding to the summands, which give nonzero
contribution.

\begin{lemma}
For the summands of the sum (\ref{<T^t_gamma>}) to be nonzero it is
necessary that

\begin{equation}
n_{i}\leq 1+\frac{t}{\left( L-P\right) },\,\,\sum_{k=1}^{P}n_{k}\geq
0
\end{equation}%
and
\begin{equation}
n_{i}\geq -1-\frac{t}{\left( L-P\right) },\,\,\sum_{k=1}^{P}n_{k}%
\leq t
\end{equation}%
if $P\leq t$ and
\begin{equation}
n_{i}\geq -1-\frac{P\left( L-P+t\right) }{L\left( L-P\right) },\,\,%
\sum_{k=1}^{P}n_{k}\leq P+\frac{t-P}{L}.
\end{equation}%
if $P>t$, for $i=1,\ldots ,P$. In the cases, when the expressions in r.h.s.
are not integer, the inequalities are strict.
\end{lemma}

From the definition of the generating function $F_{t}^{\gamma }(X,t|Y,0)$
one concludes that the coefficient of $e^{\gamma J}$ for some nonnegative
integer $J$ is the probability $P_{t}(X,J;t|Y,0;0)$ for the total distance
travelled by particles for time $t$ and the final configurations $X$,
given the initial configuration is $Y$. One can see that in (\ref%
{<T^t_gamma>}) the similar term is $e^{\sum_{i=1}^{P}\left(
x_{i}-y_{i}+Ln_{i}\right) }$, while the probability for the travelled
distance to be $J$ is the sum of the coefficients of terms, where the
sum $n_{1}+\cdots +n_{P}$ is fixed, $\sum_{i=1}^{P}\left(
x_{i}-y_{i}+Ln_{i}\right) =J$. \ Thus, the sum $(n_{1}+\cdots +n_{P})$ has a
meaning of the total number of windings around the lattice all the particles
made. By this reason, the sum is always nonnegative unlike the individual
numbers $n_{1},\ldots ,n_{P}$. The meaning of the latter is well understood
in frame of the geometric approach to the BA \cite{povolotsky priezzhev}. These are the
winding numbers of "virtual" free trajectories, which being weighted with
corresponding weights can be used to reconstruct the TASEP dynamics.

\section{\protect\bigskip Conclusion and discussion}

To conclude we have obtained the probability of the transition from one
configuration to another for arbitrary time for the TASEP with parallel
update on a ring. To this end we developed the method of summation over
the solutions of the Bethe equations, which is based on the multidimensional
version of Cauchy residue theorem. In this way the integral representation
of the solution is obtained. The expressions under the integral can be
expanded into the uniformly convergent power series, which being integrated
term by term, yields the result in form of multiple and formally infinite
sum of the terms, each having the determinant form. It is shown that only
finitely many terms of this sum are nonzero. Note that though the convergence
of the series under the integral is proved for the domain $0\leq\rho\leq1/2$, the
behaviour of the final finite sums have no singularities at the point $\rho=1/2$
as well as for any values of $\rho$.
Therefore we expect that arguments of analytic continuation exist which extend
the proof for any value of density, $0\leq\rho\leq1$. On the other hand the case
$1/2\leq\rho\leq1$ is related to $0\leq\rho\leq1/2$ by particle hole symmetry.
It is an interesting exercise to find an explicit relation between the final formulae
of the transition probabilities for these two cases.

There are several directions of possible development of the result. First, it
looks possible to generalize the method to extract not only the sum over the
solutions of the BE, but also to extract the contribution of
particular solutions. In this way one could obtain closed exact expressions
for particular eigenvalues and eigenvectors, rather than only the asymptotic
behaviour studied before. Second, the integral representation obtained can
be useful to study the large time asymptotics for the growth phenomena with time.
Many similar results
where obtained recently for the infinite lattice, due to the observed
parallels with the theory of random matrix ensembles
\cite{Johansson,rakos schutz,sas nag,Praehofer Spohn,Sasamoto,ferrari spohn1,ferrari spohn2}.
The Bethe ansatz,
giving the integral representations of the physical quantities like particle
current probability distribution, could also be a starting point of such
an asymptotical analysis. Particularly, the result of present paper could be
used to make an advance for the ring geometry where not much results have
been obtained yet.

\ack
The authors are grateful to Tony Dorlas for stimulating discussions and to
Francis Dolan for critical remarks to the text.


\bigskip

\begin{thebibliography}{99}
\bibitem{povolotsky priezzhev}
 Povolotsky A M and Priezzhev V B, 2006 J. Stat. Mech.  P07002

\bibitem{schutz} Sch\"{u}tz G M, 1997 \textit{J. Stat. Phys.} \textbf{88}
427


\bibitem{povolotsky mendes} Povolotsky A M and Mendes J F F, 2006 \textit{J.
Stat. Phys.} \textbf{123} 125

\bibitem{gwa spohn} Gwa L H and Spohn H, 1992 \textit{Phys. Rev. A} \textbf{%
46} 844

\bibitem{derrida lebowitz} Derrida B and Lebowitz J L, 1998 \textit{Phys.
Rev. Lett.} \textbf{80} 209

\bibitem{golinelli mallick}Golinelli O and Mallick K (2005) \textit{J. Phys. A: Math. Gen.} \textbf{38}  1419

\bibitem{priezzhev_preprint} Priezzhev V B \textit{Exact Non-Stationary Probabilities
in the Asymmetric Exclusion Process on a Ring} \textit{2002} Preprint cond-mat/0211052

\bibitem{priezzhev} Priezzhev V B, 2003 \textit{Phys. Rev. Lett.} \textbf{91}
050601

\bibitem{izenberg yuzhakov} Aizenberg I A and Yuzhakov A P, 1983 \textit{Integral
representations and residues in multidimensional complex analysis} (Translated
from the Russian by H. H. McFaden. Translation edited by Lev J. Leifman.
Translations of Mathematical Monographs, 58. American Mathematical Society,
Providence, RI)

\bibitem{Gaudin} Gaudin M, 1983 \textit{La Fonction d'Onde de Bethe} (Masson S. A., Paris)

\bibitem{Gaudin M McCoy B M and Wu T T} Gaudin M, McCoy B M and Wu T T 1981 \textit{Phys. Rev. D} \textbf{23} 417

\bibitem{Korepin} Korepin V E 1982 \textit{Cmmun. Math. Phys.} \textbf{86} 391

\bibitem{Johansson} Johansson K, 2000\textit{\ Comm. Math. Phys. }\textbf{\
209 }437

\bibitem{rakos schutz} R\'{a}kos A and Sch\"{u}tz G M 2005 \textit{J. Stat.
Phys.} \textbf{118} 511

\bibitem{sas nag} Nagao T and Sasamoto T, 2004 \textit{Nucl. Phys. B}
\textbf{699} 487

\bibitem{Praehofer Spohn} Praehofer M, Spohn H 2002 \textit{In and Out of
Equilibrium, edited by V. Sidoravicius,\ Progress in Probability} \textbf{51}
185

\bibitem{Sasamoto} Sasamoto T, 2005 \textit{J. Phys. A: Math. Gen. }\textbf{%
38} L549

\bibitem{ferrari spohn1} Ferrari P L and Spohn H 2006 \textit{Comm. Math.
Phys.} \textbf{265} 1

\bibitem{ferrari spohn2} Ferrari P L and Spohn H, 2005 \textit{J. Phys. A:
Math. Gen.} \textbf{38} L557


\end{thebibliography}
\end{document}